\documentclass[fleqn,usenatbib]{mnras}
\usepackage{newtxtext,newtxmath}
\usepackage{verbatim}

\usepackage[T1]{fontenc}

\DeclareRobustCommand{\VAN}[3]{#2}
\let\VANthebibliography\thebibliography
\def\thebibliography{\DeclareRobustCommand{\VAN}[3]{##3}\VANthebibliography}

\usepackage{graphicx}	
\usepackage{amsmath}	
\usepackage{subfiles}   
\usepackage{booktabs}   
\usepackage{natbib}     
\usepackage{orcidlink}
\usepackage{tabularx}
\usepackage{subfig}
\usepackage{multirow}
\usepackage{hyperref}
\usepackage{pdflscape}
\usepackage{custom-commands}

\title[The MZR and FMR at 2 < z < 8]{The \emph{JWST} EXCELS Survey: gas-phase metallicity evolution at 2 < \emph{z} < 8}

\author[T. M. Stanton et al.]{
T. M. Stanton \orcidlink{0000-0002-0827-9769},$^{1}$\thanks{E-mail: t.stanton@ed.ac.uk}
F. Cullen \orcidlink{0000-0002-3736-476X},$^{1}$
A. C. Carnall \orcidlink{0000-0002-1482-5818},$^{1}$ 
D. Scholte \orcidlink{0000-0002-6867-1244},$^{1}$ 
K. Z. Arellano-C\'ordova \orcidlink{0000-0002-2644-3518},$^{1}$ \and \
A. E. Shapley \orcidlink{0000-0003-3509-4855},$^{2}$
D. J. McLeod \orcidlink{0000-0003-4368-3326},$^{1}$ 
C. T. Donnan \orcidlink{0000-0002-7622-0208},$^{3}$  
R. Begley \orcidlink{0000-0003-0629-8074},$^{4}$ 
R. Dav\'e \orcidlink{0000-0003-2842-9434},$^{1, 5}$
J. S. Dunlop \orcidlink{0000-0002-1404-5950},$^{1}$ \and \
R. J. McLure \orcidlink{0009-0005-9742-2318},$^{1}$ 
K. Rowlands \orcidlink{0000-0001-7883-8434},$^{6, 7}$ 
C. Bondestam,$^{1}$ 
M. L. Hamadouche \orcidlink{0000-0001-6763-5551},$^{8}$ 
H.-H. Leung \orcidlink{0000-0003-0486-5178},$^{1}$ \and \ 
S. D. Stevenson \orcidlink{0000-0001-5642-752X},$^{1}$
and
E. Taylor \orcidlink{0000-0001-8728-2984}$^{1}$ 
\\
$^{1}$Institute for Astronomy, University of Edinburgh, Royal Observatory, Edinburgh, EH9 3HJ, UK\\
$^{2}$Department of Physics \& Astronomy, University of California, 430 Portola Plaza, Los Angeles CA 90095, USA\\
$^{3}$NSF’s National Optical-Infrared Astronomy Research Laboratory, 950 N. Cherry Ave., Tucson, AZ 85719, USA\\
$^{4}$Armagh Observatory and Planetarium, College Hill, Armagh, BT61 9DG, N. Ireland, UK \\
$^{5}$Department of Physics and Astronomy, University of the Western Cape, Bellville 7535, South Africa\\
$^{6}$AURA for ESA, Space Telescope Science Institute, 3700 San Martin Drive, Baltimore, MD 21218, USA \\
$^{7}$William H. Miller III Department of Physics and Astronomy, Johns Hopkins University, Baltimore, MD 21218, USA \\
$^{8}$Department of Astronomy, University of Massachusetts, Amherst, MA 01003, USA\\
}

\date{Accepted XXX. Received YYY; in original form ZZZ}

\pubyear{2026}

\begin{document}
\label{firstpage}
\pagerange{\pageref{firstpage}--\pageref{lastpage}}
\maketitle

\begin{abstract}
    We present an analysis of the gas-phase mass-metallicity relationship (MZR) and fundamental metallicity relationship (FMR) for $65$ star-forming galaxies at $2 < z < 8$ from the JWST/EXCELS survey. 
    We calculate gas-phase metallicities (12 + log(O/H)) using strong-line calibrations explicitly tested against the EXCELS sample, and report direct-method metallicities for $19$ galaxies.
    Our sample spans $8.1<\log(\rm M_\star/M_\odot)<10.3$ and $0<\log(\rm SFR/M_\odot \, yr^{-1})<2$, consistent with main-sequence star-forming galaxies at the same redshifts. 
    We find a clear MZR at both $2<z<4$ ($\langle z \rangle = 3.2$) and $4<z<8$ ($\langle z \rangle = 5.5$), with consistent slopes and mild evolution in normalization of $\simeq 0.1 \, \mathrm{dex}$, matching trends from simulations and recent observations. 
    Our results demonstrate rapid gas-phase enrichment in the early Universe; galaxies at $z \simeq 3$ (within the first $\simeq 15$ per cent of cosmic time) are enriched to $\simeq 40$ per cent of the metallicity of equivalent mass galaxies at $z=0$.
    We find tentative evidence for SFR-dependence in the MZR scatter, though results remain inconclusive and highlight the need for larger high-redshift samples. 
    Comparison with locally derived FMRs reveals a clear offset consistent with other $z > 3$ studies.
    We discuss potential drivers of this offset, noting that high-redshift samples have significantly different physical properties compared to local samples used to define the $z=0$ FMR.
    Our results confirm that low-mass, high specific star-formation rate galaxies common at high redshift are inconsistent with the equilibrium conditions underlying the local FMR, and highlight the rapid chemical enrichment at early cosmic epochs.
\end{abstract}




\section{Introduction} \label{sec:intro}

    The metallicities of galaxies are intrinsically sensitive to the overarching processes governing galaxy evolution.
    Inflows of metal-poor gas provide fuel for star formation, stars produce metals via nucleosynthesis over their lifetimes, and supernova-driven outflows remove metal-enriched gas from galaxies \citep[i.e. the baryon cycle;][]{tinsley1980, finlator&davé2008, tumlinson2017}.
    In combination with estimates of stellar masses and star-formation rates (SFRs), metallicity measurements can be considered a fundamental tracer of the evolution and growth of galaxies across cosmic time.
    Since the launch of the James Webb Space Telescope (\emph{JWST}), and in particular thanks to the unprecedented sensitivity of the Near-Infrared Spectrograph \citep[NIRSpec,][]{jakobsen2022}, there has been a significant advance in the capability of measuring accurate galaxy metallicities at high redshift ($z\gtrsim 2$).
    We are now in a position to trace chemical imprints of the cosmic baryon cycle from the epoch of reionization to the local Universe.

    One key advance made possible by \emph{JWST} has been the ability to routinely detect faint auroral emission lines (e.g. \oiiib) in galaxies between $z \simeq 2$ and $z\simeq10$ \citep[e.g.][]{arellano-córdova2022, curti2023, nakajima2023, strom2023, laseter2024, morishita2024, rogers2024, arellano-córdova2025, chakraborty2025, pollock2025}.
    Using these lines, it is possible to determine the electron temperature and derive robust gas-phase element abundances via the `gold standard' direct-$T_e$ method \citep[e.g.][]{peimbert2017, kewley2019}.
    Crucially, these direct abundance estimates can be used to construct and test new empirical strong-line metallicity calibration schemes applicable for high redshift galaxies \citep[e.g.][]{laseter2024, sanders2024, sanders2025b, cataldi2025, scholte2025}.
    Prior to \emph{JWST}, empirical strong-line calibrations were built from galaxy samples at $z<2$, meaning there was a critical systematic uncertainty when estimating the metallicities of galaxies at $z>2$, since galaxies at these redshifts are known to exhibit harder ionizing radiation fields at fixed gas-phase abundance due to their $\alpha$-enhanced massive stellar populations \citep[e.g.][]{steidel2016, topping2020a, cullen2021, kashino2022, strom2022, stanton2024, stanton2025, chartab2024, isobe2025, monty2025, rogers2025}.
    However, thanks to the new calibrations derived from \emph{JWST} data, it is now possible to robustly constrain the nebular oxygen abundances for large samples of galaxies at high redshift using the readily detectable strong lines.

    These advances have allowed the key metallicity scaling relations to be traced up to $z\simeq10$, offering constraints on the fundamental physical processes governing galaxy evolution over $> 95$ per cent of cosmic history.
    One well-known scaling relation is the relation between stellar masses and gas-phase metallicity (i.e. the mass-metallicity relationship, or MZR).
    In the MZR, galaxy metallicities increase with stellar mass following a power law before asymptotically flattening at the highest masses ($M_{\star} \gtrsim 10^{10.5} \, \mathrm{M}_{\odot}$).
    Some studies have argued that the MZR is merely a proxy of a more fundamental relationship between metallicity and gravitational potential, of which mass is a proxy, as metals are more tightly gravitationally bound in the centres of galaxies than the outskirts \citep[e.g.,][]{d'eugenio2018, sánchez-menguiano2024, boardman2025, koller2026}.
    
    The MZR has been shown to exist for both gas-phase and stellar metallicities \citep[e.g.][]{tremonti2004, gallazzi2005, andrews&martini2013, cullen2019, curti2020, sanders2021, scholte2024} and a variety of processes can be invoked to explain its shape.
    Firstly, feedback-driven outflows are more effective at removing metals from lower mass galaxies due to their shallower gravitational potential wells than higher mass galaxies \citep[e.g,][]{chisholm2018, stanton2024}.
    Higher-mass galaxies also evolve more rapidly at higher redshift than lower mass galaxies (`downsizing', \citealp{somerville&davé2015}), and as such have had more time to process their gas into metals.
    Additionally, lower-mass galaxies exhibit higher gas fractions than higher-mass galaxies \citep[e.g.][]{erb2006b}, where the on-going accretion of metal poor gas dilutes the metallicity of the ISM.
    
    The normalisation of the MZR has been observed to change with cosmic time, with observations suggesting that the MZR normalisation decreases significantly between the local universe and $z\simeq3$ \citep[decreasing by $\sim0.3\,\rm dex$;][see also \citealp{erb2006, maiolino2008, cullen2014, cullen2021, sanders2021}]{stanton2024}, then more mildly out to $z < 10$, \citep[e.g. $\sim0.1\,\rm dex$ between $z \simeq 3$ and $z \simeq 10$;][see also \citealp{nakajima2023, sanders2024}]{curti2024}, indicating that galaxies at high redshift are less oxygen enriched than their local counterparts at fixed stellar mass.
    The evolution of the slope of the MZR is subject to some debate; some studies find consistent slopes between the local Universe and cosmic noon \citep[e.g.][]{sanders2021, cullen2021, nakajima2023, stanton2024}.
    Others suggest that the MZR flattens towards higher redshifts and low stellar masses \citep[e.g.][]{li2023, curti2024, kotiwale2025}, indicating that the mechanisms regulating how inflows, outflows and stellar winds affect the metal content of the ISM may differ at lower stellar masses typical of high-redshift galaxies \citep[e.g.][]{torrey2019, scholte2024}.

    As high-redshift galaxies preferentially have lower stellar masses and higher SFRs than local main-sequence galaxies, the observed evolution in the MZR can also be explained by sampling different regions of the locally-defined redshift-invariant stellar mass-SFR-metallicity surface \citep[the fundamental metallicity relationship, or FMR; e.g.][]{ellison2008, lara-lópez2010, yates2012, mannucci2010, andrews&martini2013, cresci2019, curti2020, boardman2025}.
    This secondary dependence is substantiated by the fact that the offset of galaxies from the MZR in the local Universe (and up to $z<3$) has been shown to correlate with SFR, such that galaxies of higher SFR exhibit lower metallicities at fixed stellar mass and vice-versa.
    Accounting for this SFR dependence can effectively minimise the scatter in the MZR (e.g. $0.22\rightarrow 0.13\,\rm dex$ from the MZR to FMR, \citealp{andrews&martini2013}).
    The dependence on SFR can also be interpreted as a secondary dependence on the gas fraction; the accretion of low-metallicity or pristine gas will dilute the existing metals within the ISM whilst fuelling star formation, naturally producing an anti-correlation between the gas-phase metallicity and SFR \citep[e.g.][]{bothwell2013, lagos2016, derossi2017, scholte2024}.
    The existence of the FMR is physically motivated by gas regulator models, which describe how the gas reservoir in galaxies grows with gas accretion and decreases with star-formation and outflows \citep[e.g.][]{peeples&shankar2011, lilly2013, tacconi2020}.
    Any deviation from the FMR represents a variation in one or more of the secular processes governing galaxy evolution (i.e. star-formation and outflow efficiencies) pushing the galaxy out of an equilibrium state.
    This interpretation is supported by local galaxies which are not expected to be in equilibrium (e.g. interacting pairs) showing quantitative disagreement with the FMR \citep{maiolino&mannucci2019}.

    Gas regulator models additionally predict that the FMR should not evolve with redshift if the processes regulating galaxy evolution (i.e. inflows, star-formation and feedback-driven outflows) are themselves redshift invariant.
    Indeed, both integrated and resolved observations suggest that typical galaxies lie on the FMR over the redshift range $0 < z < 4$ \citep[e.g.][]{cresci2012, belli2013, henry2013, nakajima&ouchi2014, salim2014, hunt2016, sanders2018, arellano-córdova2024}.
    However, with the advent of \emph{JWST}, the first studies of the FMR at $z = 4 - 10$ find that galaxies preferentially exhibit metallicities lower than the local FMR would suggest \citep[e.g.][]{heintz2023, nakajima2023, curti2024, pollock2025, sarkar2025, scholte2025}, potentially signalling that the aforementioned processes do indeed differ in the early Universe. 
    The apparent deviation from the FMR is contentious as it is subject to various systematics, such as the choice of FMR parametrisation, metallicity calibration (if not direct-method), and sample selection \citep{telford2016, cresci2019}.
    For example, some studies find deviations from the local FMR at $z \sim 2$ \citep[e.g.][]{korhonen_cuestas2025} whilst others find that galaxies at $z \sim 4-6$ show no offset from the FMR \citep[e.g.][]{faisst2025a, rowland2025}.
    It is clear that large samples of galaxies with oxygen abundances measured using appropriate calibrations are required to understand the FMR and assess whether any apparent evolution is due to systematic effects or real changes in the processes governing galaxy evolution.
    
    In this work, we leverage a sample of 65 star-forming galaxies with deep rest-frame optical spectroscopy from the \emph{JWST} Early eXtragalactic Continuum and Emission Line Science survey \citep[EXCELS;][]{carnall2024}.
    Our sample spans a significant range in redshift  ($1.6 < z < 7.9$), and features a wide array of nebular emission lines from which we can estimate strong-line metallicities using calibration schemes explicitly tested on high-redshift studies such as EXCELS \citep{sanders2024, scholte2025}.
    Our sample also benefits from a substantial amount of ancillary photometric data with which we can robustly infer physical properties such as stellar masses and SFRs.
    Using these measurements we can constrain the MZR, investigate its evolution with redshift, and assess whether the FMR holds or breaks down towards higher redshifts, whilst minimising systematic differences by using a homogenous metallicity calibration and set of spectral energy distribution (SED) modelling assumptions.

    The structure of this paper is as follows.
    In Section~\ref{sec:data} we describe the EXCELS sample used in this work, and lay out our measurements of stellar masses and SFRs from nebular emission lines and SED modelling of the photometric data.
    We describe our methods to measure the strong-line and direct-method metallicities of our galaxies in Section~\ref{sec:methodology}.
    Our derived MZRs are presented in Section~\ref{sec:mzr}, alongside a discussion of systematics and the evolution of the MZR with cosmic time.
    In Section~\ref{sec:fmr} we investigate the existence and evolution of the FMR within our data, assess whether our sample shows similar offsets from the local FMR as found by recent studies in the literature, and discuss what drives any such offsets.
    Our findings and conclusions are summarised in Section~\ref{sec:conclusions}.

    Throughout, metallicities are quoted relative to a solar abundance scale taken from \citet{asplund2021}, which relates the widely-used $12 + \log {\rm (O/H)}$ to $\log (Z_{\rm gas}/Z_\odot)$ via the expression ${\rm 12 + \log(O/H)} = \log (Z_{\rm gas}/Z_\odot) + 8.69$.
    We adopt the following cosmology: $\Omega_{\rm M} = 0.3$, $\Omega_\Lambda = 0.7$, and $H_0 = 70 \, {\rm km\,s^{-1} \, Mpc^{-1}}$.

\section{Data and Sample Properties} \label{sec:data}

        \begin{figure*}
            \centering
            \includegraphics[width=0.95\linewidth]{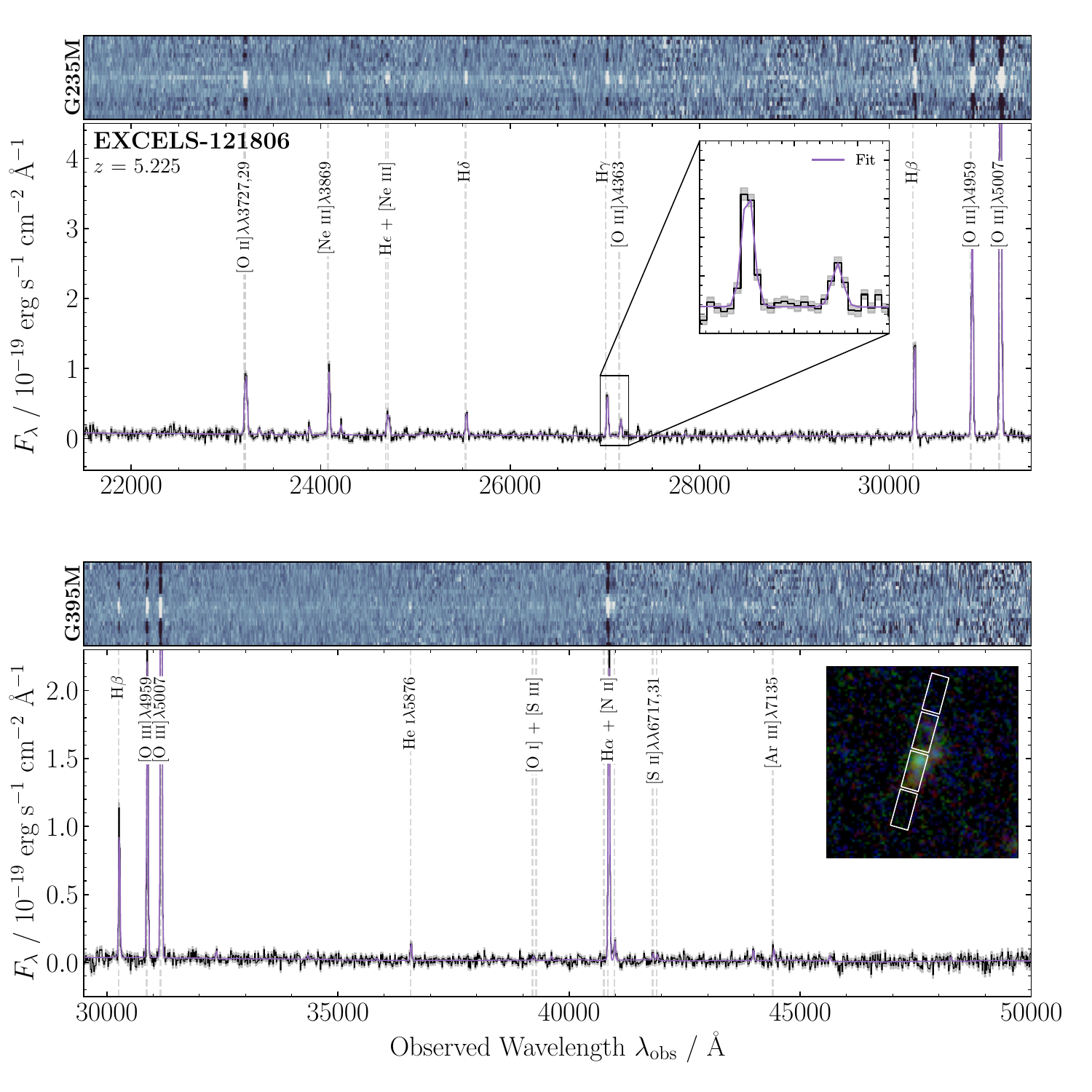}
            \caption{
            An example rest-frame optical spectrum of one galaxy in our sample at $z=5.225$ (EXCELS-121806).
            The top and lower panels show the 1D and 2D spectra for the G235M and G395M gratings respectively.
            The black lines show the 1D spectrum with the corresponding $1\sigma$ uncertainty in grey shading.
            The combined continuum and emission line model fit is plotted in purple.
            Vertical black dashed lines highlight an number of nebular emission features across the full wavelength range, with an inset in the top panel showing the \hgamma \ and \oiiib \ features.
            In the bottom panel we include a false three-colour image of EXCELS-121806 generated from the F115W + F150W, F200W + F277W, and F356W + F444W PRIMER imaging with the position of the NIRSpec MSA slitlets shown in white.
            }
            \label{fig:example-galaxy}
        \end{figure*}

    In this section, we describe the EXCELS survey, select our strong-line metallicity galaxy sample, and derive physical properties from the spectroscopic and photometric data.

    \subsection{The Early eXtragalactic Continuum and Emission Line Science survey (EXCELS)}

        The galaxy sample analysed in this work is taken from the \emph{JWST} EXCELS spectroscopic survey (GO 3543; PIs: Carnall, Cullen; \citealp{carnall2024}), which was designed to provide medium-resolution ($R=1000$) spectroscopy of high-redshift galaxy targets selected from \emph{JWST} PRIMER imaging in the UDS field (GO 1837; PI: Dunlop; \citealp{dunlop2021, mcleod2024}).
        The EXCELS observations consist of 4 NIRSpec/Micro-Shutter Array (MSA) pointings centred on rare high-redshift massive quiescent galaxies with the remaining MSA slitlets allocated to star-forming galaxies in the redshift range $1 < z < 8$.
        The target galaxies were observed using a combination of the G140M/F100LP, G235M/F170LP and G395M/F290LP gratings.
        This work is focussed on a subset of the EXCELS star-forming galaxy sample, and for which the allocated gratings provided coverage of rest-frame optical wavelengths.
        The exposure times in each grating were $\simeq 4 \, {\rm hrs}$ in G140M and G395M and $\simeq 5 \, {\rm hrs}$ in G235M.
        Full details of the sample selection and survey strategy can be found in \cite{carnall2024}.
    
        \subsubsection{Data Reduction} \label{sec:data-reduction}

        We adopt the same data reduction process and emission line measurement methodology that has been used in previous EXCELS analyses.
        In this section, we give a brief summary of these methods, but refer interested readers to \citet{carnall2024} and \citet{scholte2025} for full details.
        First, the raw level 1 data products from the Mikulski Archive for Space Telescopes (MAST) are processed using v1.17.1 of the \emph{JWST} reduction pipeline\footnote{https://github.com/spacetelescope/jwst}.
        We used the default level 1 configuration, including advanced snowball rejection, and the CRDS\_CTX = jwst\_1322.pmap version of the \textit{JWST} Calibration Reference Data System (CRDS) files.
        The level 2 and 3 pipelines are then run using the default configuration parameters to produce the combined 2D spectra.
        The final 1D science spectra are subsequently extracted using a custom optimal extraction method \citep{horne1986}, which uses the flux-weighted mean position of the object, estimated from the PRIMER imaging, within the MSA slitlet as the extraction centroid.

        We then perform a flux calibration of the 1D spectra via a two-step process.
        First, for objects observed in more that one grating, we scale the fluxes of different gratings using the median flux in the overlapping wavelength regions.
        In all cases, the G140M and/or G395M gratings are scaled to G235M.
        We then perform an absolute flux calibration to the available photometry.
        Each EXCELS galaxy benefits from PRIMER photometric coverage in the \emph{JWST}/NIRCam bands F090W, F115W, F150W, F200W, F277W, F356W, F410M and F444W.
        A subset of targets are also covered by the PRIMER \emph{JWST}/MIRI F770W imaging.
        Shorter wavelength photometry is provided by the \emph{HST} Advanced Camera for Surveys (ACS) imaging mosaics in the F435W, F606W, and F814W bands. 
        A detailed description of the photometric data is given in \citet{begley2025} (see also \citealp{mcleod2024}, McLeod et al. in prep).
        
        The \emph{HST}/ACS and \emph{JWST}/NIRcam images are PSF-homogenised to the F444W PSF and broadband fluxes are extracted in $0.5$ arcsec diameter apertures.
        The photometry is corrected to total flux using the F356W FLUX\_AUTO measurements from {\sc SourceExtractor} \citep{bertin&arnouts1996}. 
        The total flux for JWST/MIRI F770W band is extracted using $0.7$ arcsec diameter apertures.
        We integrate each spectrum through the overlapping bands and then scale the spectrum to the imaging fluxes using a linearly interpolated wavelength dependent correction between bands.
        An example spectrum for one of the star-forming galaxies analysed in this work is shown in Fig.~\ref{fig:example-galaxy}.

    \subsection{Emission line measurements} \label{sec:emission-line-measurements}
    
        The method for extracting emission line fluxes from the EXCELS data is described in detail in \citet{scholte2025}.
        Briefly, we first estimate the continuum flux using a running mean of the ${\rm 16^{th}} - {\rm 84^{th}}$ percentile flux values within a top hat function with a rest-frame width of $350\,${\AA}, which effectively masks out strong emission and absorption features.
        We then measure the emission lines from the continuum subtracted spectrum by simultaneously fitting a set of Gaussian profiles of varying amplitudes assuming a common intrinsic line width whilst allowing for instrumental broadening. 
        All lines are fit simultaneously via $\chi^2$ minimisation using {\sc scipy} \citep{virtanen2020}, and uncertainties are calculated using analytical error propagation of the line profile weighted measurements.
        
        We use the fluxes of emission lines measured simultaneously in different gratings to assess the additional flux uncertainty between gratings resulting from an imperfect flux calibration.
        We find that an average systematic uncertainty of $\simeq 8$ per cent is required to explain the observed inter-grating scatter for these lines \citep{scholte2025}.
        When calculating line ratios from lines measured from different gratings, this additional systematic uncertainty is combined in quadrature with the statistical uncertainties on the line fluxes.

    \subsection{Sample Properties and Selection} \label{sec:sample-properties}

        The full EXCELS survey comprises 340 galaxies with robust spectroscopic redshifts \citep[see][]{carnall2024}.
        Of these, 47 are quiescent galaxies that have been presented in previous analyses \citep[][]{carnall2024, skarbinski2025, stevenson2025}.
        We additionally remove three objects that show evidence of ionization by Active Galactic Nuclei (AGN) indicated by the presence of broad (i.e. $\rm FWHM>1000\,\rm km\,s^{-1}$ width) \halpha \ or \hbeta \ emission lines.
        Therefore, the parent EXCELS star-forming sample consists of 290 galaxies.

        Given the variable grating coverage of the galaxies in EXCELS, some galaxies lack spectral coverage of the necessary emission lines to estimate key parameters of the ionized gas, including dust content and metallicity.
        To select galaxies for which we can perform an accurate dust correction using the Balmer decrement, we remove any galaxy without detections of at least two Balmer lines (i.e. \halpha, \hbeta, \hgamma, or \hdelta).
        We then perform another cut to select galaxies for which we can estimate strong-line metallicities, based on detecting at least one of the following sets of lines:
        \begin{enumerate}
            \item[(i) ] \oiiia, \oii \ and \hbeta,
            \item[(ii)] \neiii, \oii \ and \hgamma.
        \end{enumerate}
        For all selections, we classify the minimum threshold for a significant line detection as $\rm S/N>3$.
        These emission-line criteria are motivated by the previous analysis of \citet{scholte2025}, who used EXCELS galaxies with direct metallicity constraints to determine optimal strong-line calibration schemes.
        We note that although metallicities can also be estimated using \nii/\halpha-based calibrations, \citet{scholte2025} found that, for the EXCELS star-forming sample as a whole, these estimates are systematically biased due to elevated N/O ratios at high redshift.
        Our final line selection ensures that we are using strong-line calibrations that are directly correlated with oxygen abundance, and that have been shown to return unbiased metallicity estimates via a comparison with direct temperature-based determinations \citep{scholte2025}.
        After applying these emission-line selection criteria, our final sample comprises $65$ galaxies for which we can robustly measure strong-line metallicities.

        \begin{figure}
            \centering
            \includegraphics[width=0.95\linewidth]{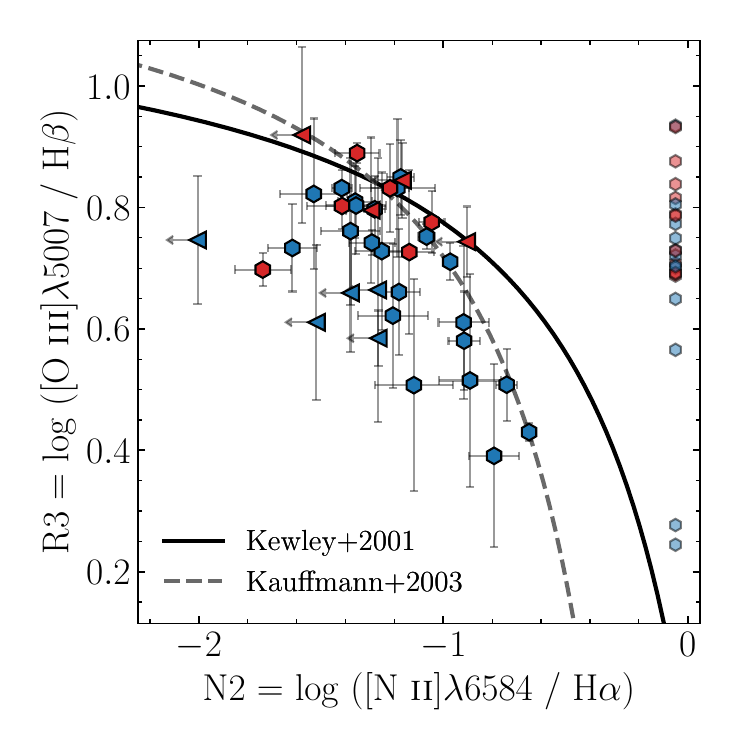}
            \caption{
            The \niinwl-BPT diagram \citep{baldwin1981} for the EXCELS galaxies in two redshift bins ($2<z<4$ in blue; $4<z<8$ in red), with $2\sigma$ upper limits shown by triangles.
            All of the galaxies in our sample with coverage of the required lines exhibit \oiiia/\hbeta \ and \nii/\halpha \ ratios fully consistent within $1\sigma$ with ionization via star-formation activity rather than AGN according to the separation lines from \citet{kewley2001} and \citet{kauffmann2003}.
            We additionally show galaxies in the selected sample which do not have coverage of \nii \ and \halpha \ as faint points at the right of the plot. 
            }
            \label{fig:nii-bpt}
        \end{figure}
        
        We perform a test for narrow-line AGN by placing our sample onto the \niinwl-BPT diagram \citep*{baldwin1981}, as shown in Fig.~\ref{fig:nii-bpt}.
        Note that, in this figure and other subsequent figures our sample is split into two redshift bins; the motivation for these two bins is given in Section \ref{sec:strong-line-mzr}.
        All of our selected EXCELS galaxies with detections of relevant lines ($28/65$, plus $9$ limits) are consistent with ionization by star formation based on the maximal starburst line defined by \citet{kewley2001}.
        As some galaxies lack spectral coverage of \nii/\halpha, we have confirmed that all our galaxies fall within the star-forming region of the mass-excitation diagram \citep{juneau2014}.
        Our sample is also consistent within $1\sigma$ with the star-forming regions of the \oiiib/\hgamma \ diagrams presented in \citet{mazzolari2024} and \citet{backhaus2025}.
        We show these diagrams in Appendix~\ref{sec:oiii4363-diagnostics}.
        Finally, none of our galaxies show evidence for AGN based on their line kinematics or feature high-ionization lines (e.g., [Ne\,{\sc iv}]$\lambda2422$, [Ne\,{\sc v}]$\lambda3420$]) typical of narrow-line AGN \citep[e.g.,][]{scholtz2025}.
        
        We note that there are additional complexities when using the classical BPT diagrams to separate star-forming galaxies from narrow-line AGN at high redshift.
        For example, some high-redshift AGN have been shown to overlap with star-forming galaxies on the \niinwl-BPT diagram \citep[][]{harikane2023, kocevski2023, maiolino2024}, whilst high-redshift star-forming galaxies show enhanced ionization through their emission line ratios (e.g., higher \oiiia/\hbeta) due to $\alpha$-enhancement \citep[e.g.,][]{shapley2024}.
        Therefore, although we cannot rule out AGN contribution to our sample with $100$ per cent confidence, we find no strong evidence for AGN contamination within our sample.
        
\begin{figure*}
            \centering
            \includegraphics[width=1.0\linewidth]{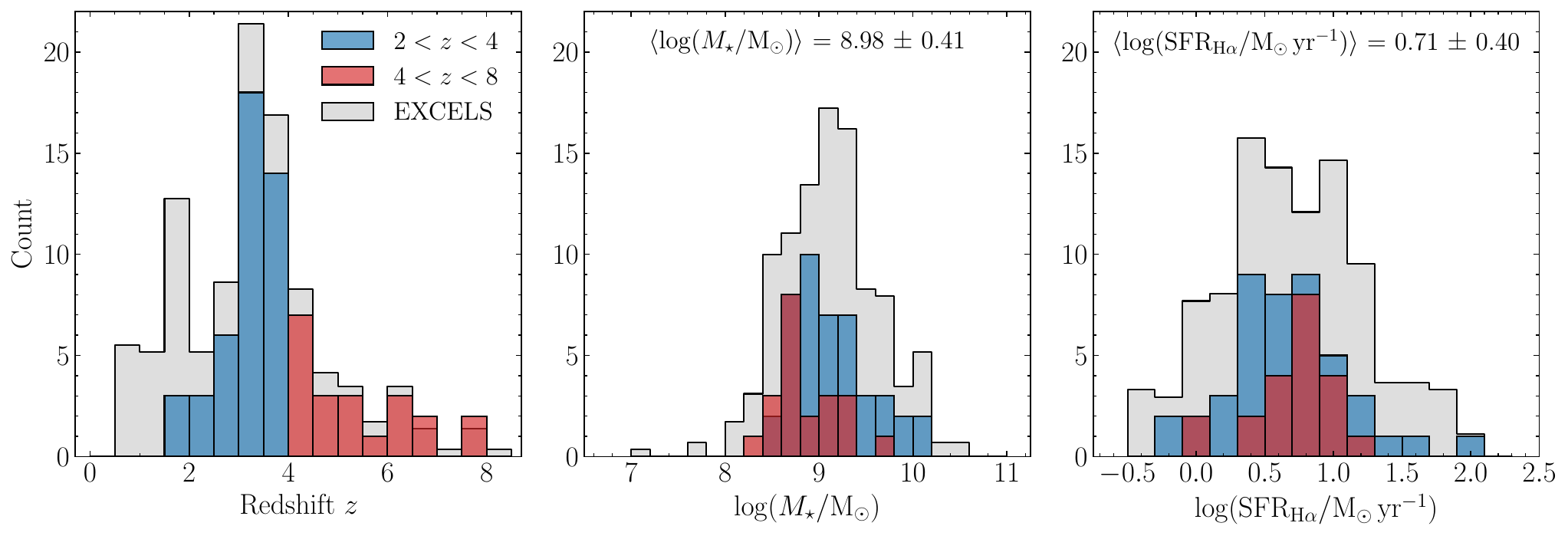}
            \caption{
            Sample distributions in $z$, $\log (M_\star / {\rm M_\odot})$ and $\log ({\rm SFR_{\rm SED}} / {\rm M_\odot \, yr^{-1}})$ for the strong-line $2<z<4$ and $4<z<8$  samples (blue and red) versus the parent sample (grey) in panels from left to right.
            For clarity, the selected sample are shown as raw counts, whilst the parent sample is shown as a percentage distribution.
            The median and standard deviation of stellar mass and SFR for the $2<z<8$ selected sample are shown at the top their respective panels.
            Our selected sample is generally representative of the EXCELS parent sample, with similar average masses of $10^{9}$ versus $10^{9.1}\,\rm M_\odot$ and star-formation rates of $5$ and $4\,\rm M_\odot\,yr^{-1}$ respectively.
            }
            \label{fig:sample_distributions}
        \end{figure*}

        Our final sample therefore consists of $65$ EXCELS star-forming galaxies in the redshift range $1.6 < z < 7.9$.
        The sample spans the stellar mass range of $8.2<\log({\rm M_\star/M_\odot})<10.2$ and a \halpha-based star-formation rate (see Section~\ref{sec:sfr-indicators}) range of $0.0<\log({\rm SFR/M_\odot\,yr^{-1}})<2.0$. 
        Details of the stellar mass and star-formation rate estimates are given in Sections \ref{sec:sed-fitting} and \ref{sec:sfr-indicators} below.
        We compare the properties of our final sample to the full EXCELS galaxy sample in Fig.~\ref{fig:sample_distributions}.

    \subsection{\textsc{Bagpipes} SED Fitting} \label{sec:sed-fitting}

        \begin{table*}
    \caption{The free parameters for our {\sc Bagpipes} SED fits to our photometric data, and their associated priors. 
    Logarithmic priors are all in base ten.}
    \label{tab:sed-fitting-parameters}
    \begin{tabularx}{\linewidth}{@{\extracolsep{\fill}}rllll@{}} 
    \toprule
    Component & Parameter & Prior Range & Prior Form & Hyperparameters\\
    \midrule \midrule
    General          & Redshift ($z$)                                          & Fixed at $z_{\rm spec}$ & $\cdots$ & $\cdots$ \\
    SFH              & Total stellar mass formed ($M_\star$ / $M_\odot$)     & ($10^5$, $ 10^{12}$)     & Logarithmic & $\cdots$ \\
    (delayed-$\tau$) & Stellar metallicity ($Z_\star$ / $Z_\odot$)           & ($0.005$, $2.0$)         & Uniform & $\cdots$ \\
                     & Time-scale ($\tau_{\rm delayed}$ / $\rm Gyr$)         & ($0.001$, $15$)          & Uniform &  $\cdots$ \\
                     & Start time ($T_0$ / $\rm Gyr$)                        & ($0.001$, $t_{\rm obs}$) & Uniform & $t_{\rm obs} = t_{\rm universe}(z)$\\
    Dust             & $V$-band attenuation ($A_V$ / ${\rm mag}$)            & ($0$, $ 6$)              & Uniform &  $\cdots$ \\
                     & Slope power law index ($\delta$)                      & ($-0.3$, $0.3$)          & Gaussian & $\mu=0.0,\,\sigma=0.1$ \\
                     & UV-bump strength  ($B$)                               & ($0$, $5$)               & Uniform &  $\cdots$ \\
    Nebular          & Ionization parameter ($\log U$)                       & ($-4$, $-1$)             & Uniform &  $\cdots$ \\
    \bottomrule
    \end{tabularx}
\end{table*}

        We derive the best-fitting spectral energy distribution (SED) for each galaxy in our sample by fitting its photometric data with the SED-fitting code {\sc Bagpipes} \citep{carnall2018, carnall2019}\footnote{When available we include the MIRI F770W band, and have verified that its inclusion does not significantly affect our best-fitting stellar masses and SED-derived SFRs.}.
        We first correct the observed photometry for nebular emission line contamination using the line fits described above. 
        The best-fitting Gaussian profiles for each detected line are used to construct a mock emission line only spectrum which is then integrated through each filter profile to determine and subtract the emission line contamination within each photometric band.
        In some cases, the bright \halpha \ line is not covered by the EXCELS data.
        For these galaxies, we estimate the \halpha \ contribution using the reddening-corrected $\hbeta$ line flux (see Section \ref{sec:dust-reddening-corrections}) and an intrinsic case-B recombination ratio of \halpha/\hbeta \ $= 2.83$ (assuming $T_e=12,000\,\rm K$, $n_e=300\,\rm cm^{-3}$; see Section~\ref{sec:dust-reddening-corrections}).

        The fitted parameters and assumed priors of our {\sc Bagpipes} modelling are outlined in Table~\ref{tab:sed-fitting-parameters}.
        We use with the Binary Population and Spectral Synthesis stellar population models \citep[BPASS v2.2.1;][]{eldridge2017, stanway&eldridge2018} assuming binary stellar populations and the default \citet{kroupa2001} IMF.
        The star-formation history (SFH) is modelled using a delayed-$\tau$ parametrisation [${\rm SFR}(t) \sim (t-T_0) \, \times \, \exp{(-(t-T_0)/\tau)}$], where $T_0$ is the time at the onset of star-formation. 
        We assume uniform priors on $\tau$ and $T_0$ of $0.01-t_{\rm universe}(z) \, \rm Gyr$ and $0.01-15\,\rm Gyr$ respectively.
        The stellar metallicity and stellar mass are varied between $0.005 \leq Z/\mathrm{Z_\odot} \leq 2$ and $10^5 \leq M_{\star}/\mathrm{M_\odot} \leq 10^{12}$ respectively, with a uniform prior on metallicity and a logarithmic prior on stellar mass.
        We model dust attenuation using the flexible attenuation law of \citet{salim2018}, assuming a flat prior on the total V-band attenuation and $2175${\AA} bump strength, and a Gaussian prior on the attenuation curve slope (see Table~\ref{tab:sed-fitting-parameters}).
        The redshift is fixed to the spectroscopic redshift for each galaxy.

        Although we have subtracted the contribution of nebular emission lines from the photometric data, the nebular continuum contribution still needs to be accounted for.
        To achieve this, we have modified the source code of {\sc Bagpipes} to include a nebular continuum component while excluding nebular emission lines. 
        We vary the strength of the nebular continuum component using the ionization parameter ($\log U$), which we include as a free parameter in our fit assuming a flat prior between $-4 \leq \log U \leq -1$.
        
        From each fit, we extract estimates and $1\sigma$ uncertainties for the stellar mass and star-formation rate (SFR) using the median and $\rm 68th$ percentile widths of their respective posterior distributions.
        For the SFR, we average over the last $10 \, \rm Myr$ of the star-formation history to give the best comparison with the \halpha-derived SFRs from the spectra (see Section \ref{sec:sfr-indicators}).
        The best-fitting stellar masses and SED-derived SFRs for our sample are given in Table~\ref{tab:sample-properties}.

        \begin{table*}
    \caption{Metallicity measurements and properties of the EXCELS sample. Metallicities are derived using the \citet{scholte2025} strong-line calibration scheme.
    }
    \label{tab:sample-properties}
    \renewcommand{\arraystretch}{1.3}
    \begin{tabularx}{\linewidth}{@{\extracolsep{\fill}}ccccccccc@{}}
    \toprule
    ID & RA & DEC & $z_{\mathrm{spec}}$ & $\log (M_\star/\mathrm{M}_\odot)$ & $\log {\rm (SFR_{SED} / {\rm M_\odot \, yr^{-1}})}$ & $\log {\rm (SFR_{H\alpha} / {\rm M_\odot \, yr^{-1}})}$ & $E(B-V)$ & $12 + {\rm \log (O/H)}$ \\ \midrule \midrule 
    40081 & $34.3659$ & $-5.2608$ & $3.955$ & $9.33^{\,+0.04}_{-0.03}$ & $1.16^{\,+0.05}_{-0.06}$ & $1.39\pm0.04$ & $0.04\pm0.03$ & $8.09^{\,+0.13}_{-0.13}$ \\
41292 & $34.2999$ & $-5.2588$ & $3.699$ & $8.86^{\,+0.05}_{-0.05}$ & $0.63^{\,+0.07}_{-0.10}$ & $0.90\pm0.06$ & $0.11\pm0.05$ & $8.00^{\,+0.17}_{-0.15}$ \\
45052 & $34.3671$ & $-5.2520$ & $4.234$ & $9.70^{\,+0.04}_{-0.04}$ & $1.53^{\,+0.05}_{-0.07}$ & $1.27\pm0.07$ & $0.38\pm0.06$ & $8.20^{\,+0.12}_{-0.14}$ \\
45177 & $34.3632$ & $-5.2523$ & $2.901$ & $8.78^{\,+0.04}_{-0.05}$ & $0.63^{\,+0.04}_{-0.06}$ & $0.64\pm0.05$ & $0.12\pm0.03$ & $8.08^{\,+0.12}_{-0.13}$ \\
45393 & $34.3538$ & $-5.2517$ & $4.236$ & $9.20^{\,+0.09}_{-0.03}$ & $1.08^{\,+0.06}_{-0.04}$ & $1.03\pm0.17$ & $0.17\pm0.11$ & $8.16^{\,+0.24}_{-0.29}$ \\
46195 & $34.2604$ & $-5.2506$ & $3.231$ & $8.94^{\,+0.07}_{-0.07}$ & $0.55^{\,+0.13}_{-0.16}$ & $-0.01\pm0.07$ & $0.07\pm0.06$ & $8.23^{\,+0.18}_{-0.20}$ \\
47143 & $34.2623$ & $-5.2489$ & $3.233$ & $8.76^{\,+0.08}_{-0.12}$ & $0.37^{\,+0.14}_{-0.13}$ & $0.46\pm0.05$ & $0.14\pm0.03$ & $8.19^{\,+0.12}_{-0.14}$ \\
47557 & $34.2621$ & $-5.2481$ & $3.234$ & $9.02^{\,+0.03}_{-0.03}$ & $0.88^{\,+0.03}_{-0.05}$ & $1.00\pm0.05$ & $0.20\pm0.03$ & $8.04^{\,+0.14}_{-0.13}$ \\
47719 & $34.3640$ & $-5.2474$ & $2.606$ & $9.66^{\,+0.04}_{-0.06}$ & $0.77^{\,+0.06}_{-0.05}$ & $1.00\pm0.08$ & $0.35\pm0.06$ & $8.46^{\,+0.10}_{-0.13}$ \\
47875 & $34.2464$ & $-5.2476$ & $3.797$ & $9.38^{\,+0.04}_{-0.05}$ & $1.24^{\,+0.05}_{-0.08}$ & $0.96\pm0.07$ & $0.17\pm0.06$ & $8.25^{\,+0.17}_{-0.23}$ \\
51029 & $34.3716$ & $-5.2424$ & $3.082$ & $9.97^{\,+0.06}_{-0.10}$ & $1.26^{\,+0.13}_{-0.09}$ & $0.49\pm0.11$ & $0.00\pm0.12$ & $8.32^{\,+0.19}_{-0.20}$ \\
52422 & $34.2503$ & $-5.2406$ & $4.025$ & $8.50^{\,+0.04}_{-0.03}$ & $0.38^{\,+0.02}_{-0.02}$ & $0.75\pm0.04$ & $0.02\pm0.03$ & $7.72^{\,+0.14}_{-0.14}$ \\
56875 & $34.3388$ & $-5.2330$ & $4.000$ & $9.03^{\,+0.03}_{-0.03}$ & $0.90^{\,+0.04}_{-0.04}$ & $0.64\pm0.04$ & $0.00\pm0.00$ & $8.17^{\,+0.12}_{-0.13}$ \\
57498 & $34.3417$ & $-5.2322$ & $3.696$ & $8.97^{\,+0.03}_{-0.04}$ & $0.84^{\,+0.04}_{-0.05}$ & $1.07\pm0.04$ & $0.14\pm0.01$ & $8.07^{\,+0.12}_{-0.13}$ \\
58103 & $34.3482$ & $-5.2308$ & $3.985$ & $9.84^{\,+0.04}_{-0.04}$ & $1.66^{\,+0.06}_{-0.09}$ & $1.53\pm0.09$ & $0.41\pm0.08$ & $8.34^{\,+0.16}_{-0.20}$ \\
58772 & $34.3046$ & $-5.2301$ & $6.109$ & $8.68^{\,+0.09}_{-0.08}$ & $0.38^{\,+0.09}_{-0.08}$ & $0.08\pm0.05$ & $0.00\pm0.04$ & $8.03^{\,+0.21}_{-0.22}$ \\
59009 & $34.3145$ & $-5.2296$ & $4.133$ & $8.73^{\,+0.05}_{-0.06}$ & $0.58^{\,+0.05}_{-0.05}$ & $0.85\pm0.05$ & $0.00\pm0.04$ & $7.97^{\,+0.14}_{-0.14}$ \\
59407 & $34.3389$ & $-5.2288$ & $3.982$ & $8.86^{\,+0.06}_{-0.05}$ & $0.57^{\,+0.09}_{-0.12}$ & $0.52\pm0.05$ & $0.05\pm0.04$ & $8.07^{\,+0.14}_{-0.14}$ \\
59720 & $34.3797$ & $-5.2283$ & $4.367$ & $9.31^{\,+0.05}_{-0.04}$ & $1.13^{\,+0.05}_{-0.06}$ & $0.81\pm0.04$ & $0.02\pm0.03$ & $8.12^{\,+0.13}_{-0.13}$ \\
60713 & $34.3135$ & $-5.2267$ & $6.171$ & $8.53^{\,+0.06}_{-0.04}$ & $0.36^{\,+0.04}_{-0.04}$ & $0.87\pm0.07$ & $0.26\pm0.06$ & $7.98^{\,+0.17}_{-0.17}$ \\
61762 & $34.2726$ & $-5.2247$ & $3.228$ & $9.29^{\,+0.15}_{-0.04}$ & $1.11^{\,+0.09}_{-0.07}$ & $0.30\pm0.04$ & $0.00\pm0.01$ & $8.15^{\,+0.12}_{-0.11}$ \\
61878 & $34.2755$ & $-5.2244$ & $3.227$ & $9.37^{\,+0.12}_{-0.14}$ & $1.03^{\,+0.10}_{-0.12}$ & $0.32\pm0.06$ & $0.05\pm0.06$ & $8.16^{\,+0.13}_{-0.16}$ \\
63629 & $34.2777$ & $-5.2213$ & $3.227$ & $8.99^{\,+0.05}_{-0.04}$ & $0.50^{\,+0.07}_{-0.06}$ & $0.60\pm0.07$ & $0.12\pm0.05$ & $8.17^{\,+0.16}_{-0.17}$ \\
63962 & $34.2809$ & $-5.2208$ & $4.359$ & $8.94^{\,+0.03}_{-0.03}$ & $0.82^{\,+0.03}_{-0.04}$ & $0.81\pm0.04$ & $0.00\pm0.01$ & $7.90^{\,+0.15}_{-0.15}$ \\
66899 & $34.3784$ & $-5.2163$ & $5.964$ & $8.25^{\,+0.05}_{-0.05}$ & $0.11^{\,+0.05}_{-0.05}$ & $0.75\pm0.06$ & $0.20\pm0.05$ & $7.64^{\,+0.18}_{-0.17}$ \\
66919 & $34.2623$ & $-5.2159$ & $3.447$ & $8.98^{\,+0.14}_{-0.06}$ & $0.71^{\,+0.09}_{-0.13}$ & $0.71\pm0.10$ & $0.42\pm0.08$ & $8.35^{\,+0.14}_{-0.16}$ \\
69991 & $34.3362$ & $-5.2110$ & $4.937$ & $8.61^{\,+0.08}_{-0.08}$ & $0.23^{\,+0.14}_{-0.12}$ & $0.47\pm0.07$ & $0.16\pm0.06$ & $8.06^{\,+0.14}_{-0.15}$ \\
70864 & $34.3496$ & $-5.2095$ & $5.255$ & $8.72^{\,+0.04}_{-0.03}$ & $0.59^{\,+0.03}_{-0.03}$ & $0.68\pm0.04$ & $0.00\pm0.00$ & $7.88^{\,+0.13}_{-0.12}$ \\
70934 & $34.3722$ & $-5.2088$ & $3.009$ & $9.74^{\,+0.03}_{-0.03}$ & $1.57^{\,+0.05}_{-0.06}$ & $1.17\pm0.05$ & $0.18\pm0.03$ & $8.37^{\,+0.10}_{-0.11}$ \\
71657 & $34.2787$ & $-5.2084$ & $6.817$ & $8.66^{\,+0.09}_{-0.10}$ & $0.40^{\,+0.10}_{-0.09}$ & $-0.01\pm0.11$ & $0.00\pm0.19$ & $8.12^{\,+0.21}_{-0.24}$ \\
73535 & $34.3523$ & $-5.2048$ & $2.210$ & $9.04^{\,+0.09}_{-0.11}$ & $0.76^{\,+0.12}_{-0.09}$ & $0.64\pm0.12$ & $0.09\pm0.07$ & $8.12^{\,+0.16}_{-0.18}$ \\
75319 & $34.3427$ & $-5.2024$ & $3.195$ & $8.43^{\,+0.12}_{-0.13}$ & $-0.04^{\,+0.12}_{-0.11}$ & $-0.26\pm0.06$ & $0.02\pm0.05$ & $8.02^{\,+0.27}_{-0.25}$ \\
75487 & $34.3395$ & $-5.2021$ & $3.241$ & $8.58^{\,+0.12}_{-0.11}$ & $0.22^{\,+0.11}_{-0.11}$ & $-0.18\pm0.04$ & $0.00\pm0.01$ & $7.92^{\,+0.25}_{-0.23}$ \\
78078 & $34.3448$ & $-5.1977$ & $3.239$ & $9.12^{\,+0.02}_{-0.02}$ & $0.99^{\,+0.03}_{-0.03}$ & $0.77\pm0.06$ & $0.18\pm0.05$ & $8.27^{\,+0.12}_{-0.12}$ \\
79373 & $34.3532$ & $-5.1957$ & $3.995$ & $9.39^{\,+0.05}_{-0.03}$ & $1.24^{\,+0.05}_{-0.05}$ & $0.90\pm0.05$ & $0.06\pm0.04$ & $8.30^{\,+0.12}_{-0.15}$ \\
83628 & $34.3401$ & $-5.1892$ & $3.464$ & $8.74^{\,+0.06}_{-0.04}$ & $0.58^{\,+0.02}_{-0.02}$ & $0.27\pm0.04$ & $0.01\pm0.02$ & $7.81^{\,+0.17}_{-0.14}$ \\
87500 & $34.2693$ & $-5.1830$ & $6.807$ & $8.64^{\,+0.04}_{-0.04}$ & $0.50^{\,+0.04}_{-0.04}$ & $0.66\pm0.05$ & $0.01\pm0.03$ & $7.83^{\,+0.17}_{-0.14}$ \\
91807 & $34.2339$ & $-5.1759$ & $3.189$ & $9.46^{\,+0.06}_{-0.08}$ & $0.93^{\,+0.17}_{-0.14}$ & $0.60\pm0.11$ & $0.29\pm0.10$ & $8.27^{\,+0.18}_{-0.22}$ \\
93404 & $34.2444$ & $-5.1736$ & $3.981$ & $8.72^{\,+0.06}_{-0.06}$ & $0.40^{\,+0.09}_{-0.08}$ & $0.47\pm0.04$ & $0.00\pm0.01$ & $8.08^{\,+0.12}_{-0.12}$ \\
93897 & $34.2633$ & $-5.1725$ & $4.080$ & $9.12^{\,+0.03}_{-0.03}$ & $0.99^{\,+0.03}_{-0.04}$ & $0.87\pm0.04$ & $0.00\pm0.02$ & $7.95^{\,+0.21}_{-0.19}$ \\
94335 & $34.2728$ & $-5.1717$ & $1.812$ & $9.37^{\,+0.12}_{-0.14}$ & $1.21^{\,+0.07}_{-0.12}$ & $0.87\pm0.04$ & $0.27\pm0.01$ & $8.18^{\,+0.11}_{-0.12}$ \\
95839 & $34.3921$ & $-5.1697$ & $4.958$ & $9.31^{\,+0.06}_{-0.06}$ & $1.09^{\,+0.06}_{-0.07}$ & $1.10\pm0.09$ & $0.12\pm0.06$ & $7.98^{\,+0.15}_{-0.15}$ \\
96066 & $34.2476$ & $-5.1692$ & $4.708$ & $9.02^{\,+0.08}_{-0.04}$ & $0.83^{\,+0.06}_{-0.08}$ & $0.61\pm0.04$ & $0.01\pm0.02$ & $8.07^{\,+0.12}_{-0.13}$ \\
96409 & $34.3857$ & $-5.1677$ & $2.293$ & $10.19^{\,+0.03}_{-0.04}$ & $2.03^{\,+0.05}_{-0.07}$ & $1.93\pm0.13$ & $0.78\pm0.09$ & $8.51^{\,+0.11}_{-0.15}$ \\
97592 & $34.3554$ & $-5.1672$ & $7.755$ & $8.93^{\,+0.09}_{-0.10}$ & $0.68^{\,+0.09}_{-0.08}$ & $0.77\pm0.07$ & $0.00\pm0.33$ & $7.87^{\,+0.17}_{-0.15}$ \\
100458 & $34.3576$ & $-5.1627$ & $3.974$ & $9.10^{\,+0.02}_{-0.02}$ & $1.00^{\,+0.02}_{-0.02}$ & $0.89\pm0.07$ & $0.15\pm0.06$ & $8.19^{\,+0.12}_{-0.14}$ \\
102088 & $34.3566$ & $-5.1601$ & $2.729$ & $9.52^{\,+0.05}_{-0.06}$ & $0.75^{\,+0.08}_{-0.07}$ & $1.28\pm0.14$ & $0.21\pm0.08$ & $8.18^{\,+0.19}_{-0.25}$ \\
103425 & $34.3864$ & $-5.1586$ & $3.063$ & $8.71^{\,+0.09}_{-0.16}$ & $0.27^{\,+0.13}_{-0.15}$ & $0.31\pm0.08$ & $0.00\pm0.39$ & $8.14^{\,+0.18}_{-0.19}$ \\
    \bottomrule
    \end{tabularx}
\end{table*}

\begin{table*}
\ContinuedFloat
    \caption{Continued.}
    \renewcommand{\arraystretch}{1.3}
    \begin{tabularx}{\linewidth}{@{\extracolsep{\fill}}ccccccccc@{}}
    \toprule
    ID & RA & DEC & $z_{\mathrm{spec}}$ & $\log (M_\star/\mathrm{M}_\odot)$ & $\log {\rm (SFR_{SED} / {\rm M_\odot \, yr^{-1}})}$ & $\log {\rm (SFR_{H\alpha} / {\rm M_\odot \, yr^{-1}})}$ & $E(B-V)$ & $12 + {\rm \log (O/H)}$ \\ \midrule \midrule 
104937 & $34.3580$ & $-5.1563$ & $1.652$ & $8.95^{\,+0.09}_{-0.14}$ & $0.05^{\,+0.11}_{-0.09}$ & $-0.03\pm0.13$ & $0.03\pm0.10$ & $8.17^{\,+0.19}_{-0.24}$ \\
105273 & $34.2319$ & $-5.1559$ & $3.185$ & $8.92^{\,+0.07}_{-0.11}$ & $0.61^{\,+0.11}_{-0.11}$ & $0.31\pm0.07$ & $0.00\pm0.50$ & $8.18^{\,+0.14}_{-0.12}$ \\
106215 & $34.2379$ & $-5.1544$ & $3.698$ & $9.48^{\,+0.08}_{-0.14}$ & $0.99^{\,+0.12}_{-0.10}$ & $0.86\pm0.07$ & $0.23\pm0.06$ & $8.21^{\,+0.13}_{-0.13}$ \\
114836 & $34.2813$ & $-5.1416$ & $5.411$ & $9.05^{\,+0.03}_{-0.03}$ & $0.92^{\,+0.03}_{-0.03}$ & $1.07\pm0.08$ & $0.18\pm0.07$ & $7.98^{\,+0.16}_{-0.17}$ \\
114961 & $34.2384$ & $-5.1412$ & $1.664$ & $9.22^{\,+0.05}_{-0.05}$ & $-0.08^{\,+0.13}_{-0.12}$ & $0.25\pm0.09$ & $0.33\pm0.08$ & $8.36^{\,+0.14}_{-0.20}$ \\
115322 & $34.4020$ & $-5.1409$ & $3.204$ & $8.68^{\,+0.17}_{-0.05}$ & $0.49^{\,+0.05}_{-0.08}$ & $0.48\pm0.06$ & $0.00\pm0.15$ & $8.08^{\,+0.15}_{-0.15}$ \\
117175 & $34.2326$ & $-5.1380$ & $6.283$ & $8.59^{\,+0.06}_{-0.06}$ & $0.40^{\,+0.05}_{-0.05}$ & $0.40\pm0.05$ & $0.00\pm0.01$ & $7.78^{\,+0.17}_{-0.16}$ \\
117563 & $34.2865$ & $-5.1372$ & $3.890$ & $9.19^{\,+0.06}_{-0.06}$ & $0.95^{\,+0.08}_{-0.12}$ & $0.64\pm0.10$ & $0.02\pm0.08$ & $8.24^{\,+0.15}_{-0.16}$ \\
118193 & $34.4008$ & $-5.1358$ & $2.205$ & $10.13^{\,+0.08}_{-0.06}$ & $1.21^{\,+0.08}_{-0.07}$ & $1.24\pm0.17$ & $0.37\pm0.10$ & $8.49^{\,+0.14}_{-0.21}$ \\
119429 & $34.2852$ & $-5.1343$ & $2.911$ & $8.96^{\,+0.05}_{-0.05}$ & $0.73^{\,+0.08}_{-0.11}$ & $0.66\pm0.11$ & $0.18\pm0.10$ & $8.04^{\,+0.23}_{-0.24}$ \\
119504 & $34.4025$ & $-5.1345$ & $7.917$ & $8.64^{\,+0.05}_{-0.05}$ & $0.51^{\,+0.04}_{-0.05}$ & $0.62\pm0.05$ & $0.00\pm0.02$ & $7.83^{\,+0.12}_{-0.12}$ \\
119533 & $34.2791$ & $-5.1340$ & $3.667$ & $8.92^{\,+0.04}_{-0.03}$ & $0.72^{\,+0.06}_{-0.07}$ & $0.44\pm0.04$ & $0.00\pm0.00$ & $8.14^{\,+0.16}_{-0.19}$ \\
120715 & $34.3633$ & $-5.1318$ & $3.077$ & $9.70^{\,+0.13}_{-0.10}$ & $1.27^{\,+0.14}_{-0.17}$ & $0.35\pm0.04$ & $0.00\pm0.08$ & $8.41^{\,+0.10}_{-0.21}$ \\
121806 & $34.4039$ & $-5.1304$ & $5.225$ & $8.75^{\,+0.04}_{-0.03}$ & $0.63^{\,+0.03}_{-0.03}$ & $0.94\pm0.05$ & $0.04\pm0.04$ & $7.90^{\,+0.15}_{-0.13}$ \\
122162 & $34.2308$ & $-5.1302$ & $2.959$ & $8.70^{\,+0.03}_{-0.03}$ & $0.57^{\,+0.04}_{-0.05}$ & $0.84\pm0.04$ & $0.00\pm0.09$ & $8.02^{\,+0.17}_{-0.16}$ \\
123597 & $34.3660$ & $-5.1280$ & $3.798$ & $9.06^{\,+0.12}_{-0.05}$ & $0.90^{\,+0.09}_{-0.05}$ & $0.84\pm0.07$ & $0.18\pm0.07$ & $8.18^{\,+0.14}_{-0.16}$ \\
123837 & $34.3817$ & $-5.1274$ & $2.617$ & $8.72^{\,+0.02}_{-0.02}$ & $0.60^{\,+0.02}_{-0.02}$ & $0.73\pm0.05$ & $0.00\pm0.04$ & $7.98^{\,+0.15}_{-0.15}$ \\
    \bottomrule
    \end{tabularx}
\end{table*}

        \subsubsection{Balmer absorption corrections} \label{sec:balmer-absorptions}

        We use to the best-fitting SED models to estimate the stellar absorption corrections for each of the Balmer lines used in our analysis (\halpha, \hbeta, \hgamma \ and \hdelta).
        To do this, we isolate each line in the best-fitting SED, subtract a linear continuum, and fit the residual absorption line with a Gaussian profile.
        To account for uncertainties, we fit 500 draws from the posterior distribution.
        We apply these correction factors to the Balmer emission line fluxes assuming an average line filling factor of 30 per cent \citep{reddy2018}.
        The average corrections to the Balmer lines (including filler factor) used in this work are $0.1$ per cent for \halpha, $0.9$ per cent for \hbeta, $2.7$ per cent for \hgamma, and $5.4$ per cent for \hdelta.

    \subsection{Dust reddening corrections} \label{sec:dust-reddening-corrections}

        A robust determination of galaxy metallicities requires correcting the observed emission line fluxes for the effects of reddening by dust.
        To infer the nebular colour excess, $E({B-V})$, we measure the ratios of Balmer line fluxes (i.e. \halpha, \hbeta, \hgamma \ and \hdelta) relative to each other and compare with their intrinsic ratios (e.g. \halpha/\hbeta \ $=2.83$, \hgamma/\hbeta \ $= 0.47$, \hdelta/\hbeta \ $=0.26$).
        The intrinsic ratios are calculated using the Python package {\sc PyNeb} \citep[v1.1.18,][]{luridiana2015, morisset+2020} assuming an average electron temperature of $T_e=12,000\,{\rm K}$ and density $n_e=300\,{\rm cm^{-3}}$.
        These values are broadly representative of the average of high-redshift galaxies in the EXCELS sample \citep{scholte2025, stanton2025}.
        We compare the intrinsic and measured values of all available line ratios, perturbing the Balmer line fluxes by their uncertainties $500$ times, and take the weighted average to determine the final $E({B-V})$ and its uncertainty.
        These values are reported for each galaxy in Table~\ref{tab:sample-properties}.
        The observed line fluxes are then corrected for reddening assuming the \citet{cardelli1989} extinction curve.

    \subsection{\texorpdfstring{H$\alpha$}{Ha}-based star-formation rates} \label{sec:sfr-indicators}

        In addition to the $10\,{\rm Myr}$ SED-based SFRs, we also estimate SFRs from the Balmer emission lines in our spectra, which are more direct probes of the recent SFR on $10\,{\rm Myr}$ timescales. 
        We first convert the dust-corrected flux density of \halpha \ into a luminosity ($L_{\rm H\alpha}$) using the spectroscopic redshift.
        For galaxies that lack spectral coverage of \halpha, we use the measured \hbeta \ flux and assume an intrinsic ratio of ${\rm H\alpha}/{\rm H\beta} = 2.83$.
        We then convert $L_{\rm H\alpha}$ to a star-formation rate using the \citet{clarke2024} conversion which is suitable for high-redshift galaxies with sub-solar metallicities:
        \begin{equation} \label{eq:ha-sfr-calibration}
            {\rm \log (SFR_{H\alpha} / M_\odot\,yr^{-1})} = -41.59 + \log_{10}\left( \frac{L_{\rm H\alpha}}{\rm erg/s}\right).
        \end{equation}
        
        We also test the recent SFR calibration introduced by \citet{korhonen_cuestas2025} in which the conversion factor (the constant in equation \ref{eq:ha-sfr-calibration}) is stellar metallicity dependent, $C(Z_\star)$, and ranges from $-41.7$ to $-41.4$ across the range $Z_\star/\mathrm{Z_{\odot}} \simeq 0.05 - 1.0$.
        For the ionizing spectra of stars, the relevant element abundance is iron (Fe), which is in deficit relative to oxygen in high-redshift star-forming galaxies \citep[e.g.][see also \citealp{méndez-delgado2024}]{cullen2021, stanton2024}.
        We convert the measured oxygen abundances of our galaxies (see Section \ref{sec:strong-line-metallicities}) to iron abundances based on the average O/Fe enhancement at $z\simeq3.5$ derived by \citet{stanton2024} and apply the appropriate conversion constant based on the estimated iron abundance.
        The resulting SFRs differ from the \citet{clarke2024} estimates by $-0.09<\Delta\log({\rm SFR/M_\odot\,yr^{-1}})<0.02\,\rm dex$, with an average of $-0.05\,\rm dex$.
        These offsets are well within the statistical uncertainty on any individual SFR measurement.
        For the remainder of this analysis we use the SFRs assuming the uniform \citet{clarke2024} correction, but note that applying a metallicity dependent correction would not alter the conclusions of this work.
        
\begin{figure*} 
            \centering
            \includegraphics[width=\linewidth]{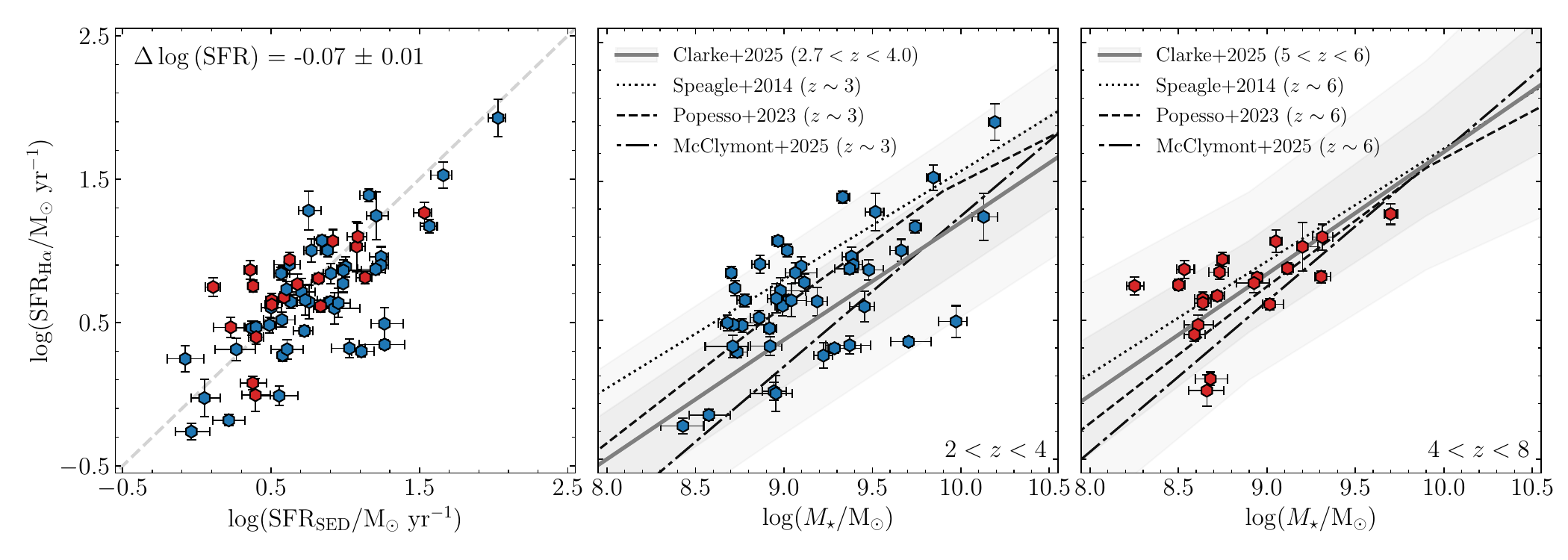}
            \caption{
            In the left panel, we show the agreement between our SED- and \halpha-based SFR probes (with $2<z<4$ galaxies in blue and $4<z<8$ galaxies in red).
            We find an average offset of $-0.07\pm 0.01\,{\rm dex}$, indicating that our SED modelled parameters are representative of our spectroscopically derived parameters.
            In the centre and right panels we show the  ${\rm SFR_{H\alpha}}$ star-forming main sequence for our sample in two redshift bins: $2 < z < 4$ (blue hexagons) and $4 < z < 8$ (red hexagons).
            For each redshift bin we plot a series of empirical relations at similar redshifts from \citet[][black dotted line]{speagle2014}, \citet[][black dashed line]{popesso2023}, \citet[][grey line and shaded region]{clarke2025}, and the theoretical prescription from \citet[][black dot-dashed line]{mcclymont2025a}.
            In both the $2 < z < 4$ and $4 < z < 8$ redshift bins $94$ per cent of galaxies fall within at least $2\sigma$ of the star-forming main sequence, indicating these galaxies are broadly representative of the general star-forming galaxy populations at these epochs.
            }
            \label{fig:star-forming-main-sequence}
        \end{figure*}

        In the left panel of Fig.~\ref{fig:star-forming-main-sequence} we show a comparison between our Balmer line-based SFR estimates (which we refer to as  ${\rm SFR_{H\alpha}}$ for simplicity) and the SED-based estimates.
        It can be seen that, in general, both estimates are highly consistent.
        On average, the ${\rm SFR_{H\alpha}}$ values are offset to lower SFR by $-0.07 \, \rm dex$ ($\simeq 15$ per cent) but this is not enough to affect any of the conclusions presented here.
        Overall, Fig.~\ref{fig:star-forming-main-sequence} demonstrates a reassuring self-consistency between our SED-derived and spectroscopically-derived physical parameters.
        For the remainder of this work, we adopt ${\rm SFR_{H\alpha}}$ as our best estimate of the true recent SFR.

    \subsection{The star-forming main sequence} \label{sec:the-sfms}

        To assess how representative our sample is of the general star-forming galaxy population at $z>2$ we compare with various literature estimates of the star-forming main sequence (SFMS) at $2<z<8$.
        In the centre and right-hand panels of Fig.~\ref{fig:star-forming-main-sequence} we show the SFMS for $2 < z < 4$, and $4 < z < 8$, respectively.
        In each redshift window, we show the observational SFMS parametrisations of \citet{speagle2014}, \citet{popesso2023}, and \citet{clarke2025} and the theoretical SFMS presented in  \citet{mcclymont2025a}.
        Overall, we find excellent agreement with the \citet{clarke2025} SFMS with $65$ per cent of our total sample falling within their estimated $1\sigma$ scatter, and $94$ per cent within the $2\sigma$ scatter.

        Accounting for the differences in slope and normalisation across the different SFMS prescriptions we find that, averaged across these independent estimates, our sample clearly falls within expectations of the SFMS.
        Whilst our sample at $2<z<4$ shows a marginal offset towards higher SFR with respect to the \citet{clarke2025} SFMS, given the scatter in our data there is with no strong evidence for a significant bias towards more highly star-forming galaxies.
        Based on this comparison, our EXCELS sample - although not complete at any stellar mass - can be considered to be broadly representative of the typical star-forming galaxy population at these epochs.

\section{Measuring Metallicities} \label{sec:methodology}

    The primary aim of this study to characterise the mass-metallicity and fundamental metallicity relations in our EXCELS $2<z<8$ sample using strong-line diagnostics. 
    However, a subset of this strong-line sample also benefits from \oiiib \ auroral line detections that can be used to estimate direct-method abundances \citep[e.g.,][]{scholte2025}.
    Throughout this analysis we also make use of this auroral line sample, primarily for sanity-checking the results derived from the strong lines.
    Below, we describe the methodologies for calculating metallicities in both cases.

    \subsection{Strong-line Metallicities}\label{sec:strong-line-metallicities}

        Our observations cover a range of nebular emission lines in the rest-frame optical regime associated with multiple individual metal species (i.e. O, Ne, Ar, S, N).
        Whilst line ratios involving N, Ar and S are useful tracers of oxygen abundance on average \citep[e.g.][]{stasińska2006, bian2018, curti2020, sanders2025b}, these elements have different enrichment pathways to O that can in principle bias O/H estimates for individual galaxies (for example in N-enhanced systems; see discussion in \citealp{scholte2025}).
        We therefore restrict our analysis to the \oii, \neiii, \hgamma, \hbeta, and \oiii \ emission lines (see Section \ref{sec:sample-properties}) and the following strong-line metallicity diagnostics:
        
        \begin{itemize}
            \item R3 = \oiiia\,/\,\hbeta,
            \item R2 = \oii\,/\,\hbeta,
            \item O32 = \oiii\,/\,\oii,
            \item Ne3O2 = \neiii\,/\,\oii.
        \end{itemize}

        For these diagnostic ratios, and others, the conversion to oxygen abundance has been shown to vary with redshift due to the evolving ionization conditions in the local to high-redshift Universe \citep[e.g.][]{sanders2024, sanders2025b}.
        Calibrations derived using galaxies in the local Universe are therefore typically not appropriate for high-redshift galaxies (unless the ionization conditions are matched; e.g. \citealp{bian2018}).
        Recently, \citet{laseter2024} introduced the ${\rm \hat{R}} = 0.47{\rm R2} + 0.88{\rm R3}$ diagnostic, an optimal rotation of the $\rm R2$--$\rm R3$--$12+\log(\rm O/H)$ plane to minimise the effect of the evolution in ionization parameter and enable application to galaxies across all redshifts.
        This diagnostic was recently re-calibrated by \citet{scholte2025} using a large sample galaxies from the Dark Energy Spectroscopic Instrument (DESI) Early Data Release (EDR).
        \citet{scholte2025} additionally introduced the ${\rm \hat{R}_{Ne}} = 0.47{\rm R2} + 0.88{\rm Ne3}$ diagnostic, a variation of $\rm \hat{R}$ where R2 = \oii/\hgamma \ and Ne3 = \neiii/\hgamma, which can be used when the \oiiia \ line is not available.
        Crucially, these calibrations have been shown to accurately recover the metallicities of EXCELS galaxies with direct temperature-based abundance measurement (\citealp{scholte2025}; see also Section~\ref{sec:direct-method-metallicities}).
        They have also been shown to work well for galaxies up to $z\simeq 10$ \citep[e.g.][]{donnan2025}.
        
        To estimate the metallicities we minimise the following $\chi^2$ equation:
        \begin{equation} \label{eqn:strong-line-chi-squared}
            \chi^2 (x) = \sum_i \frac{({\rm R_{obs,\,i}} - {\rm R_{cal,\,i}}(x))^2}{({\rm \sigma^2_{obs,\,i}} + {\rm \sigma^2_{cal,\,i}})},
        \end{equation}
        where we sum over each available line ratio ($i$), $x = 12 + \log (\rm O/H) - 8.69$, ${\rm R_{obs,\,i}}$ is the logarithm of the $i$th line ratio, ${\rm R_{cal,\,i}}$ is the predicted value of the logarithm of the line ratio at $x$, and $\sigma_{\rm obs,\,i}$ and $\sigma_{\rm cal,\,i}$ are the uncertainties on the measured line ratio and calibration respectively.
        
        Both the ${\rm \hat{R}}$ and ${\rm \hat{R}_{Ne}}$ calibrations turnover/plateau at $12+\log(\rm O/H) \simeq 8.0$.
        To break the degeneracy between the upper and lower branches we implement an additional monotonic calibration from \citet{sanders2024}, specifically $\rm O32$ for ${\rm \hat{R}}$ and $\rm Ne3O2$ for ${\rm \hat{R}_{Ne}}$.
        We convert equation~\ref{eqn:strong-line-chi-squared} into a likelihood function ($\propto \exp(-\chi^2/2)$) which we then maximise over $x$ using the nested sampling package {\sc dynesty} \citep{speagle2020, koposov2023}.
        We set a flat metallicity prior of $6.7 \leq 12+\log(\rm O/H) \leq 9.0$, ($0.01-2\,\rm Z_\odot$), which is a slight extrapolation of calibration range in \citet{scholte2025} ($6.9 \leq 12 + \log(\rm O/H) \leq 8.6$).
        However, none of our derived metallicities fall outside of this original calibration range.
        We are able to fit $39/65$ of our galaxies with both the $\rm \hat R$ and  $\rm \hat R_{Ne}$ calibrations, whilst $21/65$ galaxies and $5/65$ galaxies have constraints from only $\rm \hat R$ and  $\rm \hat R_{Ne}$ respectively.
        The final metallicity is calculated from the median of the resulting posterior distribution in $x$, and the $1\sigma$ uncertainty is given by the 68$^{\rm th}$ percentile width of the posterior distribution.
        When fitting each galaxy, we exclude any ratio with an emission line detection below a $3\sigma$ detection threshold.

    \subsection{Direct-method Metallicities} \label{sec:direct-method-metallicities}

        Following the methodology outlined in \citet{cullen2025} and \citet{scholte2025}, we adopt a forward modelling approach to estimate the direct method abundance for the $19/65$ galaxies in our sample with a ${\rm SNR} > 3$ detection of \oiiib.
        
        Our model has five free parameters: the high-ionization zone electron temperature ($T_e\,$[{\sc O iii}]), the electron density ($n_e$), the $V$-band dust attenuation ($A_V$) and the abundances of singly- and doubly-ionized oxygen, $\log (\rm O^+/H^+)$ and $\log (\rm O^{++}/H^+)$ respectively.
        To explore the parameter space, we use the nested sampling package {\sc dynesty}, and adopt the following uniform priors.
        The electron temperature and density are permitted to vary across the ranges $3\leq\log(T_e/{\rm K})\leq5$ and $1\leq\log(n_e/{\rm cm^{-3}})\leq4$ respectively, the abundances of $\log (\rm O^+/H^+)$ and $\log (\rm O^{++}/H^+)$ from $-8$ to $0$, and the $V$-band attenuation from $0$ to $4\,\rm magnitudes$.
        The best-fitting model for each galaxy is constrained using a set of observed emission-line ratios, including dust-sensitive ratios (e.g. \halpha/\hbeta \ or \hgamma/\hbeta), temperature-sensitive ratios (i.e. \oiiib/\oiiia), abundance-sensitive ratios (e.g. \oiiia/\hbeta, \oii/\hbeta) and, if available, density-sensitive ratios (e.g. \siia/\siib).
        For galaxies that do not have coverage of the \sii \ doublet we instead marginalise over the prior range of densities.
        We do not expect this to bias our abundance measurements since the density does not strongly affect our adopted line ratios unless $n_e \gtrsim 3500\,{\rm cm^{-3}}$, which is higher then the typical density of typical $2 < z < 8$ galaxies \citep[$100-1000\,\rm cm^{-3}$, e.g.,][]{isobe2023, méndez-delgado2023, topping2025}.
        
        To estimate the low-ionization zone temperature ($T_e\,$[{\sc O ii}]) from $T_e\,$[{\sc O iii}] we assume the \citet{campbell1986} relationship:
        \begin{equation} \label{eq:temperature-scaling}
            T_e[{\rm O} \ \textsc{ii}] = 0.7 \times T_e[{\rm O} \ \textsc{iii}] + 3000.
        \end{equation}
        This relationship has been shown to be appropriate for systems below $T_e\,$[O {\sc iii}]$\, \lesssim20,000\,\rm K$ \citep[][]{arellano-córdova&rodriguez2020, rogers2021, sanders2025b, scholte2026}.
        Some JWST studies have suggested a shallower relationship between such that $T_e([{\rm O} \ \textsc{ii}])$ would be lower than suggested by equation~\ref{eq:temperature-scaling} at $T_e([{\rm O} \ \textsc{iii}])>20,000\,\rm K$ \citep[e.g.][]{cataldi2025, chakraborty2025}.
        However, temperature relationships are not well constrained in the high-temperature regime due to small numbers of galaxies with $T_e([{\rm O} \ \textsc{iii}])>20,000\,\rm K$ available to calibrate against.
        Additionally, $T_e\,$[O {\sc ii}] measurements from \oii \ and \auroraloii \ are subject to further dispersion due to the additional dependence on density \citep{méndez-delgado2023}.
        Nonetheless, the majority of our sample has $T_e([{\rm O} \ \textsc{iii}]) \, < 20,000\,\rm K$ where both local and high-redshift calibrations are in good agreement.

        We report the electron temperatures and direct-method metallicities for the 19 galaxies in our auroral-line sample, along with \sii \ density constraints for $7$ of the $19$ galaxies, in Table~\ref{tab:direct-metallicities}.
        The direct-method and strong-line metallicities for these $19$ galaxies show good agreement, with a minimal offset of $-0.08\pm0.05\,\rm dex$, within the typical uncertainties of either metallicity measurement. 
        Comparing our results with \citet{scholte2025}, we find excellent agreement for the objects in common, with an average offset of $0.01\pm0.04\,\rm dex$.
        We discuss the comparison with \citet{scholte2025} further in Appendix~\ref{sec:direct-method-appendix}.

\section{The Evolution of the Mass-Metallicity Relationship} \label{sec:mzr}

    \begin{figure*}
        \centering
        \includegraphics[width=1.0\linewidth]{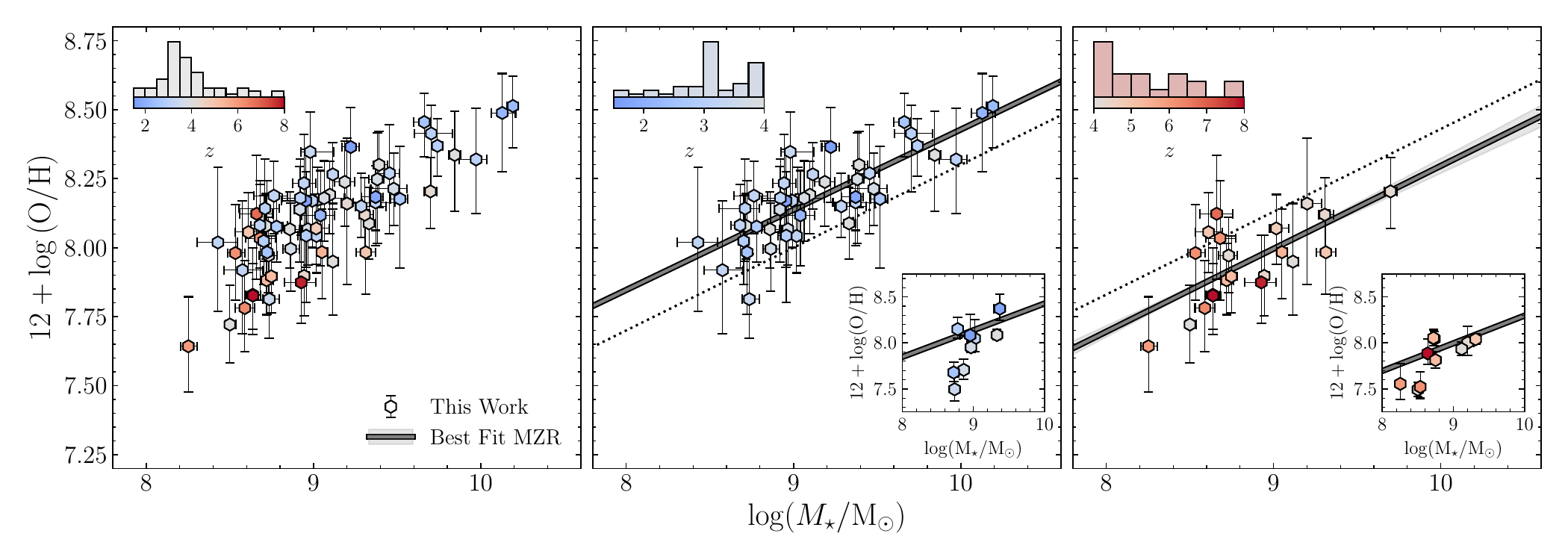}
        \caption{
        The strong-line metallicity based MZR for the EXCELS sample (hexagons) using the \citet{scholte2025} calibrations for the full sample ($\langle z \rangle = 4.0$, left panel), $2 < z < 4$ ($\langle z \rangle = 3.2$, centre panel) and $4 < z < 8$ ($\langle z \rangle = 5.5$, right panel).
        Each point is colour coded according to its redshift, with the associated colour bar in the top left of each panel, above which is plotted a histogram showing the redshift distribution of each sample.
        The best fit MZR and $1\sigma$ uncertainty for the $2<z<4$ and $4<z<8$ subsamples are shown by solid grey lines and a grey shaded region in the centre and right panels respectively, with the alternate redshift MZR shown as a black dotted line for comparison.
        In the centre and right panels, we additionally show inset panels with the direct-method metallicities plotted against the strong-line MZR in each redshift regime.
        }
        \label{fig:mass-metallicity-relationship}
    \end{figure*}

    In this section, we present the MZR derived from the strong-line metallicities of our galaxies, and assess robustness of the strong-line determination using the direct-method metallicities.
    We then investigate whether there is evidence for evolution in the MZR across the redshift range of our sample, and compare our results with recent observational and simulation-based determinations.
    
    \subsection{The Mass-Metallicity Relationship} \label{sec:strong-line-mzr}

        In the left panel of Fig.~\ref{fig:mass-metallicity-relationship} we show the best-fitting metallicities of the full EXCELS sample as a function of stellar mass.
        Our sample exhibits a clear trend between $12+\log {\rm (O/H)}$ and $M_\star$, such that gas-phase metallicity increases with increasing stellar mass.
        By colour-coding our galaxies by redshift we see a clear distinction between the low- ($2<z<4$) and high-redshift ($4<z<8$) regimes such that, at fixed mass, high-redshift galaxies have preferentially lower metallicities than their lower redshift counterparts.
        Motivated by this split, we create two subsamples of galaxies by dividing our sample at the average redshift of $\langle z \rangle=4$, and show individual mass-metallicity relationships for these two bins in the centre and right panels of Fig.~\ref{fig:mass-metallicity-relationship}.
        
        To both our $2<z<4$ and $4<z<8$ subsamples, we fit the logarithm of the gas-phase metallicity as a linear function of the logarithm mass (i.e. the MZR) using the following form:        
        \begin{equation}\label{eqn:general-mzr}
            12 + \log {\rm(O/H)} = \beta \, m_{9} + Z_{9},
        \end{equation}
        where $m_{9} = \log (M_\star/{\rm 10^{9} \, M_\odot})$.
        To perform the fit we employ an orthogonal linear regression accounting for errors in both stellar mass and metallicity following the methodology described in \citet{hogg2010}.

        In the $2<z<4$ redshift bin (with an average redshift of $\langle z \rangle\simeq3.2$), we find the following best-fitting relationship:
        \begin{equation}\label{eqn:2z4-best-fit-mzr}
            12 + \log {\rm(O/H)} = (0.29 \pm 0.01) \, m_{9} + (8.14 \pm 0.01).
        \end{equation}
        The slope of our fit is in broad agreement with the low-mass slope (i.e. at $M_\star < 10^{10}\,\rm M_\odot$) of the $z=0$ MZR from \citet{scholte2024} of $0.24\,\rm dex$, with a normalisation offset of $\simeq0.35\,\rm dex$ implying a factor $\simeq 2.2$ decrease in $\rm O/H$ between $z=3.2$ and $z=0$. 
        Our results indicate that at $z\simeq3.2$ the oxygen abundance varies from $12 + \log(\rm O/H)=7.85-8.50$ (i.e. $0.14-0.65\,\rm Z_\odot$) across a stellar mass range of $M_\star = 10^8-10^{10.25}\,\rm M_\odot$, with an average abundance of $8.14\pm0.01$ ($\simeq0.28\,\rm Z_\odot$) at $M_\star = 10^9\,\rm M_\odot$.

        In the $4 < z < 8$ redshift bin (with an average redshift of $\langle z \rangle=5.5$), the best-fitting relation has the form:
        \begin{equation}\label{eqn:4z8-best-fit-mzr}
            12 + \log {\rm(O/H)} = (0.30 \pm 0.02) \, m_{9} + (8.00 \pm 0.03).
        \end{equation}
        The slope is fully consistent with the lower redshift bin with an evolution in the normalisation of $\simeq -0.1\,\rm dex$.
        We discuss this evolution in more detail in Section~\ref{sec:mzr-evolution}.
        The offset from the low-mass $z=0$ MZR of \citet{scholte2024} is $\simeq0.5\,\rm dex$, indicating a factor $\simeq3.1$ decrease in $\rm O/H$ from $z=0-5.5$.
        The metallicities implied by our best fit relationship at $z\sim5.5$ suggest that the oxygen abundance varies from $12 + \log(\rm O/H)=7.70-8.38$ (i.e. $0.10-0.47\,\rm Z_\odot$) across a stellar mass range of $M_\star = 10^8-10^{10.25} \, \rm M_\odot$, with an average abundance of $8.00\pm0.03$ ($\simeq0.20\,\rm Z_\odot$) at $M_\star = 10^9\,\rm M_\odot$.

        Taken together, our results suggest that the metal enrichment of the Universe is initially rapid, with galaxies at fixed mass enriched - on average - to $\simeq 30$ per cent of the oxygen abundance of galaxies at $z\simeq0$ by $z \simeq 5.5$ (i.e. in the first $1\,\rm Gyr$ of the Universe's history).
        Over the next billion years enrichment progresses more slowly, with galaxies at $\simeq 40$ per cent of their present day metallicity by $z=3.2$.
        We discuss this trend further in Section~\ref{sec:cosmic-metallicity-evolution}.

    \subsubsection{Comparing $T_e$-based metallicities to the strong-line MZR} \label{sec:strong-line-direct-comparison}

        Due to the small size of our direct-method metallicity sample (25 galaxies) and, at least in the lower-redshift bin, its narrow range in stellar mass, it is difficult to infer a MZR using these galaxies alone.
        However, it is instructive to use this sample to assess the robustness of the results derived from our strong-line analysis.
        The inset panels of Fig. \ref{fig:mass-metallicity-relationship} show the direct-method metallicities compared to the strong-line MZRs in each redshift bin.
        
        This comparison makes it immediately clear that: (i) the direct-method metallicities are generally consistent with the strong-line derived MZRs and (ii) in the $z < 4$ bin, where the direct-method sample occupies a narrow range in stellar mass, there is evidence that the strong-line analysis underestimates the true MZR scatter.
        The galaxies at $z<4$ with direct-method metallicities show and average offset of $\simeq 0.14\,\rm dex$ below the strong-line $2<z<4$ MZR.
        This is mainly driven by the three galaxies with $12+\mathrm{log(O/H)} < 7.8$, and is likely a consequence of the bias towards lower metallicities when measuring direct method metallicities (due to the increasing strength of \oiiib \ with decreasing metallicity).
        At $z>4$ there is a much less significant offset to lower metallicities of $\simeq 0.07\,\rm dex$, potentially due to the $4 < z< 8$ redshift range containing lower metallicity galaxies on average, and the \oiiib \ emission line shifting to wavelengths where \emph{JWST}/NIRSpec is more sensitive.
        Overall, however, the comparison is favourable, and suggests the MZR at high redshifts can be derived from strong-line analyses as long as reasonable empirical calibrations are used. 
        
        That being said, there is evidence that, in both redshift bins, the strong-line metallicities are less scattered around the MZR compared to the direct method metallicities.
        In the $2<z<4$ bin, the strong-line metallicities exhibit a scatter of $\sigma=0.09\,\rm dex$ compared with $0.28\,\rm dex$ for the direct-method metallicities.
        The difference is less stark at $4<z<8$, with the strong-line and direct-method metallicities exhibiting scatter of $\sigma\sim0.10\,\rm dex$ and $\sim0.17\,\rm dex$ respectively.
        Therefore, these results indicate that strong-line calibrations may be artificially reducing the intrinsic scatter in the MZR by a factor $\simeq 2-3$.
        This reduction in the scatter of abundances derived from strong-line methods has been discussed previously in the context of the MZR \citep[e.g.][]{steidel2014} and abundance relations \citep[e.g.][]{vale-asari-2016} and is fundamentally linked to the variation in ionization conditions at fixed metallicity that is not accounted for in the majority of empirical strong-line calibrations.
        For example, for the \citet{scholte2025} calibrations employed here, the scatter in ${\rm O/H}$ at fixed $\hat{\rm R}$ is $\sigma_{\rm \hat{R}}=0.13\,\rm dex$, which is \emph{smaller} than the scatter in our derived MZR. 
        Clearly, the observed scatter for our strong-line sample is simply a reflection of the scatter in the calibration, and does not capture the intrinsic scatter of the relation.
        Although these calibrations will yield accurate average metallicities, the artificial reduction in the scatter is worth noting.
        Reproduction of the true scatter of metallicity around the MZR requires the inclusion of ionization parameters into metallicity calibrations, which as of yet few empirical calibration schemes implicitly include \citep[e.g.][for calibrations built from photoionization models see \citealp{kk04_calibration, blanc2015}]{nakajima2022}.
        
    \subsection{Comparison with the literature and simulations} \label{sec:mzr-lit-comparison}

        \begin{figure*}
            \centering
            \includegraphics[width=\linewidth]{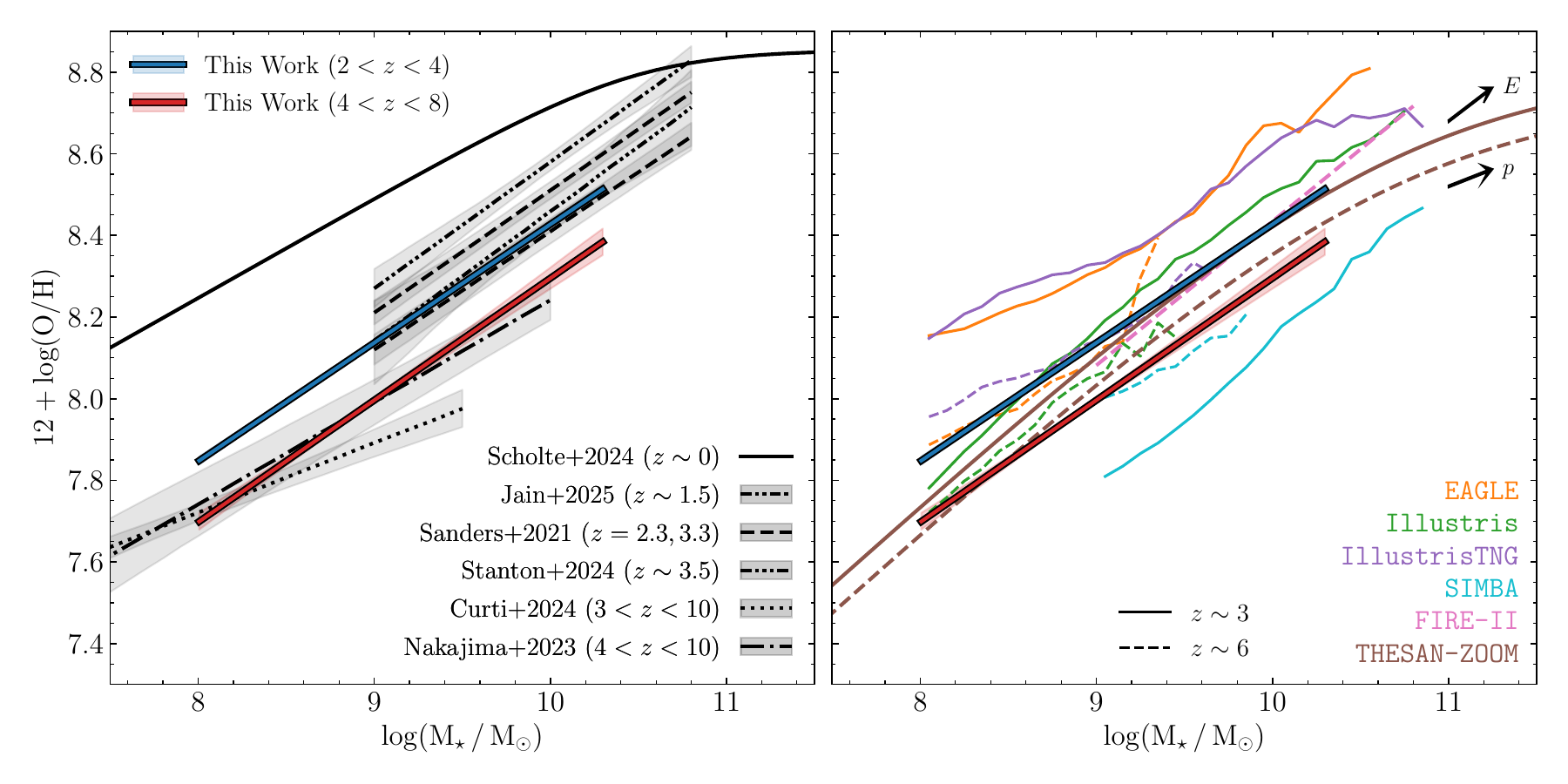}
            \caption{
            A comparison of the MZRs derived in this work at $2 < z < 4$ ($z=3.2$, blue line), and $4 < z < 8$ ($z=5.5$, red line), with determinations from the literature.
            In the left panel, we show observationally derived MZRs across a variety of epochs, including the $z\sim0$ relationship derived by \citet{scholte2024} using galaxies from DESI (black solid line), the $z\simeq1.5$ relationship (black double-dot dashed line) from \citet{jain2025}, the $z=2.3,\,3.3$ results from the MOSDEF survey \citep[black dashed lines;][]{sanders2021}, the $z\simeq3.5$ relationship from the NIRVANDELS survey \citep[black triple-dot-dashed line][]{stanton2024}, the $3 < z < 10$ relationship from JADES \citep[black dotted line;][]{curti2024}, and the determination from CEERS galaxies at $4 < z < 10$ \citep[black dot-dashed line;][]{nakajima2023}.
            We see excellent consistency with the relationships of \citet{sanders2021} and \citet{stanton2024} for our $2 < z < 4$ relationship in terms of slope and normalisation.
            At $4 < z < 8$ we see broad agreement with with the \citet{nakajima2023} relation for our $4 < z < 10$ relationship, though our slope is marginally steeper.
            Our $4 < z < 8$ slope is significantly steeper than the relationship of \citet{curti2024}, though their data extends to lower stellar masses where some studies suggest the gradient of the MZR flattens.
            In the right-hand panel, we show our relationships compared to the results from simulations at $z\sim3$ (solid lines) and $z\sim6$ (dotted lines).
            We plot the MZRs for the {\sc Illustris} (green lines), {\sc IllustrisTNG} (purple lines), {\sc SIMBA} (light blue lines) and {\sc EAGLE} (orange lines) simulations, as compiled by \citet{garcia2025}.
            Parametrisations of the MZR from {\sc Fire-ii} \citep{marszewski2024} and {\sc Thesan-zoom} \citep{mcclymont2025b} are plotted in pink and brown respectively.
            We additionally plot the predicted slope of the MZR for energy- ($E$) and momentum-driven ($p$) outflows from the semi-analytical models of \citet{guo2016} in the top right corner.
            }
            \label{fig:mzr-lit-sim-comparison}
        \end{figure*}

        In this sub-section we place our derived MZRs into a broader context via a comparison with other independent literature determinations (left-hand panel of Fig. \ref{fig:mzr-lit-sim-comparison}) and the predictions of cosmological simulations and semi-analytic models (right-hand panel of Fig. \ref{fig:mzr-lit-sim-comparison}).
        Where necessary we convert the reported normalisation of literature MZRs to their values at $10^{9}\,\rm M_\odot$.
        
        Uniformly across all relationships, both observed and simulated, the MZR evolves towards lower metallicities with increasing cosmic time, and shows a fairly consistent consistent power law slope at stellar masses $ < 10^{10} \, \rm M_\odot$ at all redshifts.
        We find excellent agreement between our MZR fit at $z \simeq 3.2$ and the $z=3.3$ ground-based MZR from \citet{sanders2021}, who report low-mass slope of $0.29\pm 0.02$ and normalisation of $8.12\pm 0.04$, fully consistent within $1\sigma$ of our relationship. 
        We also note similar agreement with the results of \citet{stanton2024} at $z=3.5$ from the NIRVANDELS survey.
        For our $z > 4$ bin, we find broad agreement between our relationship and the results of \citet{nakajima2023}, who find a normalisation of $7.99\pm0.06$, and a slope consistent within $2 \sigma$ ($0.25\pm0.03$).
        Additionally, we find good agreement with the $4<z<10$ MZR of \citet{sarkar2025}, who recover a slope and normalisation of $0.27\pm0.02$ and $8.01\pm0.08$, respectively.
        The \citet{curti2024} MZR exhibits a significantly flatter MZR, with a gradient of $0.18\pm0.03$.
        However, the \citet{curti2024} sample probes a lower stellar masses compared our sample ($M_{\star} \sim 10^7 - 10^9 \, \mathrm{M}_{\odot}$), a stellar mass regime where other studies have also previously reported a flattening in the MZR \citep[e.g.][]{li2023}.
        Our data does not show any evidence for a flattening at $M_{\star} < 10^9 \, \mathrm{M}_{\odot}$ (Fig. \ref{fig:mass-metallicity-relationship}) but we ultimately lack the numbers and the dynamic range in stellar mass to draw any firm conclusions.
        On the other hand, our MZR at $z>4$ is also shallower than some other high-redshift determinations, for example \citet{heintz2023} ($7<z<10$; $\beta=0.33$) and \citet{rowland2025} ($6 < z< 8$; $\beta=0.37\pm0.03$).
        However, we stress that one must take care when comparing MZRs derived using differing strong-line calibrations, and from samples with differing selection functions.
        What we can conclude is that our methodology, using strong-line method proven to recover direct-method metallicities at $z > 2$, suggests no evolution in the slope of the MZR between $z=0$ and $z\simeq6$ (in the stellar mass range $M_{\star} \sim 10^{8.5} - 10^{10.5} \, \mathrm{M}_{\odot}$).

        \begin{figure*}
            \centering
            \includegraphics[width=\linewidth]{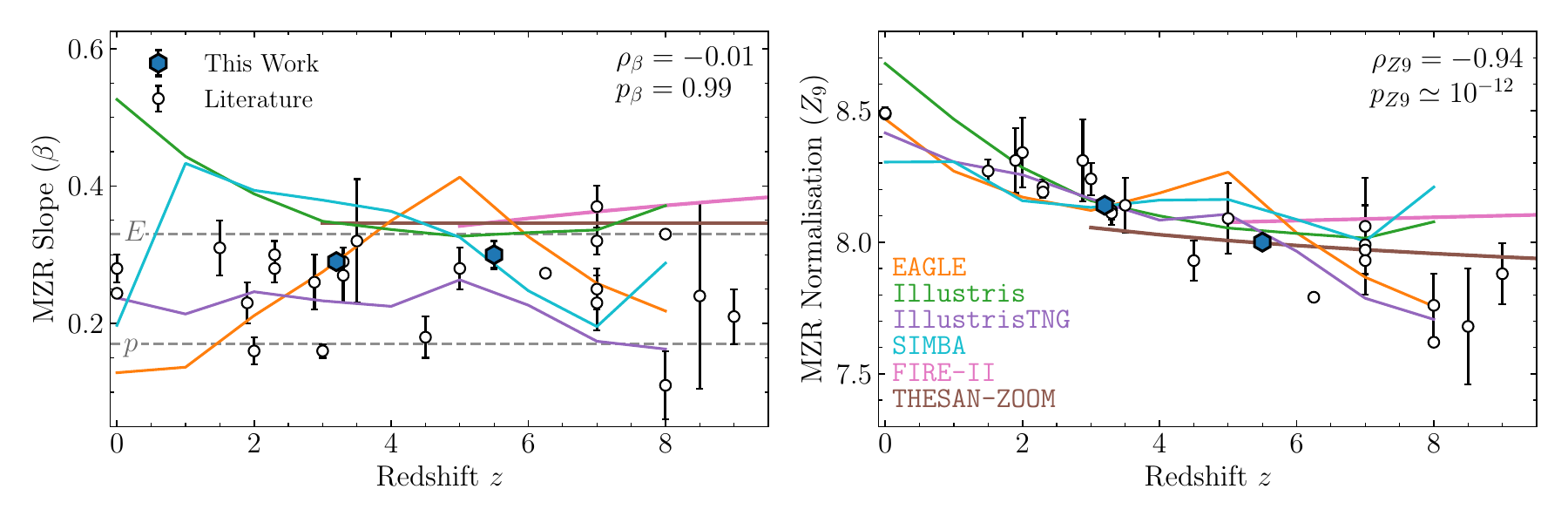}
            \caption{
            The redshift evolution of the slope ($\beta$, left panel) and normalisation ($Z_{9}$, right panel) of the MZR for this work (blue), a compilation of studies from the literature (grey), and cosmological simulations as detailed in the text.
            The slope of the MZR shows no clear sign of evolution with cosmic time, being in place as early as $z\sim10$ (i.e. $500\,\rm Myr$ after the Big Bang), with a Spearman Rank correlation coefficient of $\rho=-0.01$ and a $p$-value of $0.99$ indicating no significant correlation between slope and redshift.
            The normalisation however shows a clear decreasing trend with redshift, decreasing by $\sim1\,\rm dex$ over $0 < z < 10$. 
            A Spearman rank analysis finds $\rho=-0.94$ with a $p$-value of $\sim10^{-12}$ indicating a statistically significant negative correlation.
            We additionally show the results of cosmological and zoom-in simulations, adopting the same colours as in Fig.~\ref{fig:mzr-lit-sim-comparison}.
            In the left panel, we show the slopes corresponding to energy- and momentum-driven winds with grey dashed lines.
            The simulations exhibit a variety of slopes, likely due to differences in their underlying physics, with no clear sign of an evolving slope with redshift.
            In the right panel we re-scale all of the normalisations to match our measurement at $z\sim3.2$, finding a clear consensus of a decreasing normalisation of the MZR with increasing redshift.
            }
            \label{fig:mzr-parameter-evolution}
        \end{figure*}

        In the right-hand panel of Fig.~\ref{fig:mzr-lit-sim-comparison} we compare our results to the predictions of various galaxy-formation simulations.
        We plot the average metallicity as a function of stellar mass at $z=3$ and $z=6$ from four cosmological simulations, {\sc EAGLE} \citep{crain2015, schaye2015}, {\sc Illustris} \citep[][]{vogelsberger2014, nelson2015}, {\sc IllustrisTNG} \citep[][]{pillepich2018, marinacci2018, naiman2018, nelson2018, nelson2019, springel2018}, and {\sc SIMBA} \citep{davé2019}, as compiled by \citet{garcia2025}.
        Additionally, we plot parametrisations of the MZR at $z=3$ and $z=6$ from the {\sc Thesan-Zoom} \citep{mcclymont2025b} simulations, and at $z\sim6$ from the {\sc Fire-II} \citep{marszewski2024, marszewski2025} simulations.
        The average metallicity increases with increasing stellar mass in all of the simulations, and qualitatively shows similar gradients to our observations, although the normalisation of the MZR varies, often substantially (e.g. up to $\simeq0.4\,\rm dex$), between simulations.
        This is perhaps unsurprising, as each simulation utilises different implementations of the underlying physics regulating star formation and metallicity evolution -  including stellar and AGN feedback\footnote{Notably, the difference between the $z=3-6$ {\sc SIMBA} MZRs appears to be inverted, in contrast to our observations and the results of other simulations.
        This is an artefact of the implementation of feedback in {\sc SIMBA}, wherein the mass-loading factor of outflows is attenuated at high redshift to boost early star formation, but with the resulting side-effect of yielding higher metallicities at higher redshift than at low redshift \citep[][]{finlator&davé2008, jones2024}.} - and different assumed supernovae yields \citep[as discussed in][]{garcia2024a, garcia2024b}.
        However, in terms of relative metallicity evolution at fixed stellar mass between $z\simeq6$ and $z\simeq3$, the various simulations generally reproduce the evolution we observe (i.e. in that redshift interval the various simulations predicted an increase in log(O/H) of ${0.1 - 0.2 \, \rm{dex}}$, compared to the observed $\simeq 0.13 \, \rm{dex}$).

        It can be seen from Fig. \ref{fig:mzr-lit-sim-comparison} that the slope of the MZR is also relatively consistent across all simulations and similar to the observed value.
        The MZR slope can be considered to be regulated - to first order - by feedback-driven outflows, and therefore is indicative of the mode of feedback.
        Semi-analytic modelling, and gas-regulator models \citep[e.g.][]{davé2012, peeples&shankar2011, lilly2013, tacconi2018}, have been used to test the effect of applying different modes of feedback on the resulting slope of the MZR.
        In Fig. \ref{fig:mzr-lit-sim-comparison} we show the predicted gradients the energy-driven ($\beta\simeq0.33$) and momentum-driven ($\beta \simeq 0.17$) wind prescriptions, as estimated by \citet{guo2016} using the equilibrium models of \citet{davé2012}.
        Our derived MZRs show better alignment with the steeper slope predicted for energy-driven winds.
        The momentum-drive wind prescription aligns well with the shallower MZR of \citet{curti2024} (as well as \citet{li2023} and \citet{kotiwale2025}).
        However, as mentioned above, those MZRs apply to lower mass galaxies (i.e. $10^7-10^9\,\rm M_\odot$).
        These results suggest that the feedback mechanisms regulating the slope of the MZR change may change around at transition mass of $\simeq 10^9 \, \mathrm{M}_{\odot}$.
        Interestingly, \citet{scholte2024} have reported a similar transition in the MZR slope at $M_{\star} \lesssim 10^9 \, \mathrm{M}_{\odot}$ at $z \simeq 0$ using a mass-complete sample of galaxies from the DESI survey.
        Using atomic gas measurements for their sample, \citet{scholte2024} argued that the transition to a shallower slope at low mass was directly related to the shape and slope transition of the atomic gas sequence.
        Currently, a similar transition in the high-redshift MZR slope at $\simeq 10^9 \, \mathrm{M}_{\odot}$ can be inferred via a comparison between different studies; however, a significantly larger, and homogenously analysed, sample panning multiple decades in stellar masses is still needed to definitely confirm this trend.

    \subsection{The redshift evolution of the MZR} \label{sec:mzr-evolution}

        In this sub-section we situate our new MZR determinations at $z\simeq3.2$ and $z\simeq5.5$ within the broader context of the MZR evolution between $z=0$ and $z\simeq10$.
        In Fig.~\ref{fig:mzr-parameter-evolution} we show the evolution of the slope ($\beta$) and normalisation ($Z_{9}$) of the mass-metallicity relationship reported in this work and across a range of studies in the literature. 
        The various literature studies have unique sample selections and adopt a number of different strong-line calibration schemes \citep{sanders2021, heintz2023, langeroodi2023, li2023, nakajima2023, chemerynska2024, curti2024, he2024, scholte2024, stanton2024, jain2025, rowland2025}.
        We also include the MZR determination from \citet{morishita2024} who use direct-method metallicities only.
        Where required, we have adjusted the reported MZR parameters to be centred around the same stellar mass (i.e. the normalisation is taken at $10^{9}\,\rm M_\odot$).
        
        As these different fits span a wide range of assumptions (i.e. calibration methodology) and contain individual selection functions (e.g. the stellar mass ranges of their samples), we caution that the discussion below should be considered qualitatively, and that larger homogenously analysed samples are required to quantitively asses any redshift trends in these parameters.
        Nevertheless, a number of interesting trends are evident in Fig.~\ref{fig:mzr-parameter-evolution}. 
        First, although the individual literature measurements of the MZR slope exhibit some diversity, the current picture is consistent with no evolution with redshift.
        The majority of slope measurements fall between the regimes specified for momentum- and energy-driven winds (i.e. $\beta=0.17-0.33$), with most more closely aligned with steeper energy-driven scenario. 
        A Spearman rank analysis of the trend between slope and redshift indicates no correlation ($\rho=-0.01, \,p=0.99$), suggesting that the slope of the MZR is in place as early as $z \simeq 10$.
        We note that we are not accounting here for the possible change in slope at $M_{\star} < 10^9 \, \mathrm{M}_{\odot}$ and the various mass regimes sampled in the different studies.
        However, it seems clear that most studies find a similar MZR slope below the high-metallicity plateau (i.e. at $M_{\star} \lesssim 10^{10.5} \, \mathrm{M}_{\odot}$).
        
        In contrast, the normalisation shows - as expected - a clear negative trend with redshift, such that the average metallicity at $\log (\rm M_\star/M_\odot)=9$ increases by up to $\simeq1\,\rm dex$ between $z=10$ and $z=0$.
        Performing a similar Spearman rank analysis yields $\rho=-0.94, \, p\sim10^{-12}$, indicating a strong correlation between redshift and the normalisation of the MZR.
        Considering just the MZR at $z=0$ and our new results at $z=3.2$ and $z=5.5$, we find that ${\rm d\log (O/H)}/{\rm d}z=-0.09\pm0.01$.
        Including all the literature data in Fig.~\ref{fig:mzr-parameter-evolution}, we find a fully consistent value of ${\rm d\log (O/H)}/{\rm d}z=-0.10\pm0.01$.
        We explore the full time-evolution of galaxy metallicities further in Sec~\ref{sec:cosmic-metallicity-evolution} below.
        
        In Fig.~\ref{fig:mzr-parameter-evolution} we also compare to the evolution of the slope and normalisation of the MZR for the simulations presented in the previous section.
        For this, we use the parametrisations presented in \citet{mcclymont2025b} for {\sc Thesan-zoom} and \citet{marszewski2025} for {\sc FIRE-II}.
        For the {\sc EAGLE}, {\sc Illustris}, {\sc IllustrisTNG} and {\sc SIMBA} simulations we fit an MZR of the form of equation~\ref{eqn:general-mzr} to all galaxies within the stellar mass range $8 \leq \log(\rm M_\star /M_\odot) \leq 10.5$ as a function of redshift.
        Since the simulations show significant differences in their normalisations (Fig.~\ref{fig:mzr-lit-sim-comparison}), we instead show the relative evolution in normalisation for each simulation by scaling each to match our measurement at $z=3.2$.
        
        The slopes of the MZR presented across simulations generally vary between $0.2 <\beta < 0.4$, which generally aligns well with our compilation of observationally-derived slopes.
        Similar to the observations, there is no clear consensus on how the MZR slope evolves with redshift.
        The {\sc FIRE-II}, {\sc Thesan-zoom} and {\sc IllustrisTNG} simulations show relatively little evolution with redshift whereas both {\sc EAGLE} and {\sc SIMBA} show an initial rise before turning over and decreasing towards higher redshift.
        However, the trends seen between the MZR normalisation and redshift show a much more consistent picture. 
        All simulations predict similar relative metallicity evolution in agreement with literature measurements up to $z\simeq4$.
        Interestingly, at $z>4$, the simulations begin to diverge further motivating the need for more accurate constraints on the MZR at these redshifts.
        
        Considering both slope and normalisation, the {\sc IllustrisTNG} simulations show the best agreement with the observational measurements implying the inflow and outflow processes regulating the baryon cycle are more appropriately approximated in {\sc IllustrisTNG} compared to other simulations (see also \citealp{jain2025}).
        Overall, despite some uncertainty in the true slope of the MZR, the consistent trend between the normalisation and redshift across both observations and simulations is reassuring.

        \subsubsection{Comparison to the \citet{jain2025} evolving MZR parametrisation} \label{sec:evolving-mzr}

            \begin{figure}
                \centering
                \includegraphics[width=\linewidth]{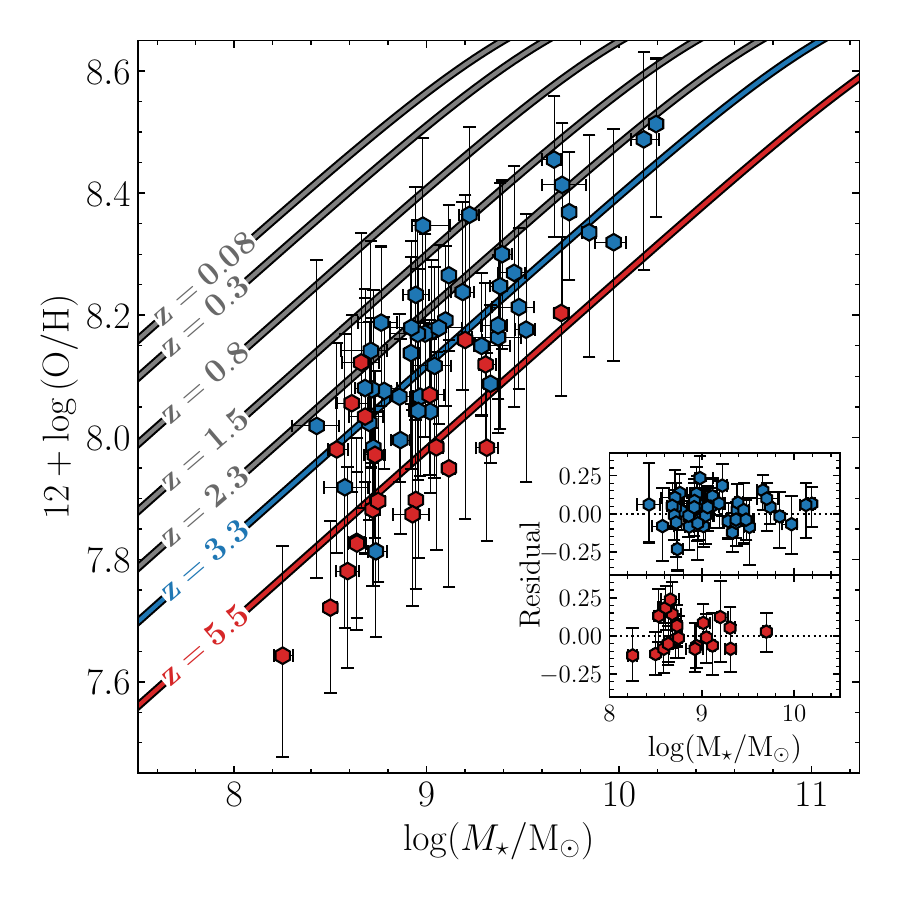}
                \caption{
                The EXCELS galaxies, at $2 <z < 4$ (blue) and $4 < z < 8$ (red), alongside the redshift-evolving MZR parametrisation from \citet{sanders2023} and \citet{jain2025} (grey lines). 
                We extrapolate the MZR from $z=3.3$ out to our high-redshift bin at $z=5.5$.
                Our data appears evenly scattered around the MZR for both redshift bins, as further highlighted in the inset panels showing the residual metallicity for each bin.
                }
                \label{fig:evolving-mzr}
            \end{figure}

            Evidence for a similar MZR shape at all redshifts has led some authors explore redshift-dependent parametrisations of the MZR.
            For example, \citet{sanders2023} proposed the following relation with a fixed low-mass slope and redshift-dependent turnover mass:
            \begin{equation} \label{eq:evolving-mzr}
                {\rm 12 + log(O/H)}({\rm M_\star}, z) = {\rm Z_0} - \frac{\beta}{\Delta} \times \log \left( 1 + \left(\frac{\rm M_\star}{{\rm M_{0}}(z)} \right)^{-\Delta} \right),
            \end{equation}
            where $\beta$ is the low-mass slope of the MZR, $Z_0=8.76$ is the high-mass asymptotic metallicity, ${\rm M_0}(z) = m_1 + m_2 \log (1+z)$ is the turnover mass, and $\Delta=1.2$ acts to smooth the transition between the power-law and the turnover.
            \citet{sanders2023} calibrated this relationship using a compilation of stacked measurements at $z=0.08$ \citep{curti2020}, $z=0.8$ \citep{zahid2011}, $z=1.5$ \citep{topping2021}, $z=2.3$ and $z=3.3$ \citep{sanders2021}.
            The relationship was recently recalibrated by \citet{jain2025} using additional data at $z=0.3$ and $z=1.5$.

            Motivated by the excellent quantitative agreement between our measured MZR slope and normalisation at $z=3.2$ with these studies (Fig.~\ref{fig:evolving-mzr}), we extrapolate the parametrisation of \citet{jain2025} to $z = 5.5$ and compare with our data.
            As expected we find that galaxies in our low-redshift bin ($2<z<4$; $z\sim3.2$) show excellent agreement with their parametrisation, with a average offset consistent with zero ($0.04 \pm 0.09 \,\rm dex$).
            At $z = 5.5$, we also see excellent agreement between our measurements and the extrapolation of the \citet{jain2025} relation, with an average offset of $-0.01 \pm 0.10 \,\rm dex$.
            Our results therefore suggest that the \citet{jain2025} parametrisation extends up to at least $z\simeq6$.
            With larger samples at $z>6$ becoming more readily available with subsequent \emph{JWST} cycles, it will  soon be possible to assess whether this relationship holds into the epoch of reionization.

    \subsection{The rapid enrichment of galaxies at early epochs} \label{sec:cosmic-metallicity-evolution}

        In this sub-section, motivated by our agreement with the parametrisation of \citet{jain2025}, we consider an alternative view of the evolution of the MZR by presenting the metallicity of galaxies with $10^9 \, \rm{M}_{\odot}$ across cosmic time as a fraction of the metallicity of an equivalent mass galaxy at $z\simeq0$ (i.e $\mathrm{12+log(O/H)}= 8.56$ from \citealp{jain2025}), which we define as $f_{\rm O/H}$.

        \begin{figure}
            \centering
            \includegraphics[width=\linewidth]{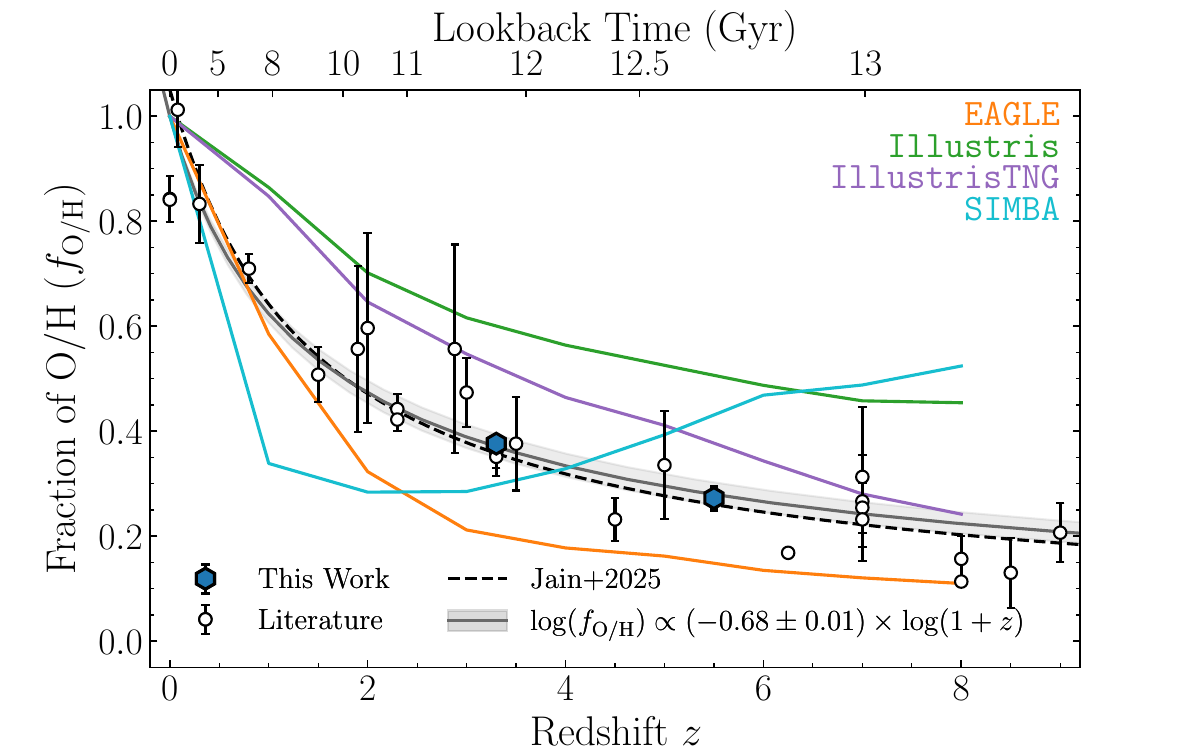}
            \caption{
            An alternative representation of the evolution of the MZR showing the fraction of O/H ($f_{\rm O/H}$) calculated from our MZR fits (outlined blue hexagons) and MZR fits from the literature (black circles), with respect to the local MZR evaluated at $10^9\,\rm M_\odot$.
            We plot the trends from {\sc EAGLE} (orange line), {\sc Illustris} (green line), {\sc IllustrisTNG} (purple line) and {\sc SIMBA} (light blue line), where $f_{\rm O/H}$ has been estimated assuming the $z=0$ MZR in each simulation.
            Our results, and the results from observational and simulated MZRs, indicate a power-law relationship between $f_{\rm O/H}$ and redshift.
            We fit a simple illustrative model to the MZR data, plotted in black with $3\sigma$ uncertainties in grey.
            }
            \label{fig:cosmic-metallicity-evolution}
        \end{figure}
        
        In Fig.~\ref{fig:cosmic-metallicity-evolution} we plot $f_{\rm O/H}$ as a function of redshift based the MZRs presented here (evaluated at a fixed mass of $10^9\,\rm M_\odot$).
        It is important to note that this does not represent the metallicity evolution of individual galaxies at $10^9\,\rm M_\odot$ over cosmic time (i.e. it is not an evolutionary sequence for  $10^9\,\rm M_\odot$ galaxies).
        Instead, Fig.~\ref{fig:cosmic-metallicity-evolution} is an alternative way of presenting of the evolving normalisation of the MZR, and aims to contextualise how enriched galaxies at $10^9\,\rm M_\odot$ are at different cosmic epochs.
        
        As discussed previously, we find that the fraction of metallicity with respect to $z=0$ galaxies increases from $\simeq 30$ per cent at $z\simeq5.5$ to $\simeq 40$ per cent at $z \simeq 3.2$.
        Combined with the $f_{\rm O/H}$ values calculated from our compilation of literature MZRs, we find that the redshift evolution is best parametrised with a power law as $\log(f_{\rm O/H}) \propto \log(1+z)$ with
        \begin{equation} \label{eq:metallicity-fraction}
            \log(f_{\rm O/H})=(-0.68\pm0.01) \times \log (1 + z),
        \end{equation}
        when fixing $f_{\rm O/H}=1$ at $z=0$.
        Our model agrees with the general trend extracted from the \citet{jain2025} relationship, though begins to plateau at $z>8$.
        This model illustrates that typical $10^{9}\,\rm M_\odot$ galaxies are already enriched at $\simeq20$ per cent of the metallicity of equivalent mass galaxies at $z=0$ by $z\simeq10$, $\simeq30$ per cent by $z\simeq6$, and $\simeq40$ per cent enriched by $z\simeq3$.
        These values highlight the rapid metal enrichment of the galaxy population within the first $\simeq 2-3\,\rm Gyr$, followed by a more gradual enrichment across the following $\simeq 10 \, \rm Gyr$.
        Further constraints on galaxies metallicities at $z>8$ are necessary to determine the relative enrichment at cosmic dawn more precisely.

        In Fig.~\ref{fig:cosmic-metallicity-evolution} we again compare with the predictions of cosmological simulations by extracting the metallicity evolution of $M_{\star} = 10^9 \, \mathrm{M}_{\odot}$ galaxies within each simulation.
        It can be seen that the general shape of the enrichment history is similar to the observations across all four simulations, with some notable differences. 
        The behaviour of {\sc SIMBA} shows a remarkably rapid increase in metallicity fraction across between $z=2$ and $z=0$, and a decreasing metallicity fraction from $z=8$ and $z=2$. 
        This is again due to the implementation of feedback at the highest redshifts in SIMBA (as discussed in Section~\ref{sec:mzr-lit-comparison} and \citealp{jones2024}).
        For example, {\sc EAGLE} predicts a much more gradual increase in metallicity up to $z\simeq2$ but, at $z<1$, provides the closest match to the compiled data points and implied trend from our illustrative model.
        {\sc Illustris} and {\sc IllustrisTNG} predict similarly gradual increases in metallicity across $2<z<8$, although at a much higher fraction of present day metallicity.
        Towards higher redshifts we note better agreement with the {\sc IllustrisTNG} simulation.
        However, no simulation completely reproduces the trends in typical metallicity of $10^{9}\,\rm M_\odot$ galaxies, or MZR normalisation at fixed mass.
        This indicates that, while there is still much work to be done observationally in constraining the MZR, there is still room to improve the chemical evolution across cosmic time within simulations to better match current observed metallicities \citep[see also][]{khostovan2025}.

\section{The Fundamental Metallicity Relationship} \label{sec:fmr}

    In this section, we investigate whether there is evidence for a SFR dependence in the scatter of galaxy metallicities at fixed stellar mass.
    SFR-dependent scatter is present in the $z=0$ MZR for galaxies with $M_{\star} > 10^9 \, \mathrm{M}_{\odot}$ and is driven in part by the anti-correlation between metallicity and SFR at fixed mass referred to as the `fundamental metallicity relation' (FMR; \citealp{mannucci2010}).
    We then assess whether our sample falls onto the locally-defined FMR and compare with similar studies from the literature and the predictions of cosmological simulations.

    \subsection{Evidence for an FMR in our EXCELS sample via parametric and non-parametric tests} \label{sec:strong-line-fmr-empirical}

        \begin{figure*}
            \centering
            \includegraphics[width=0.95\linewidth]{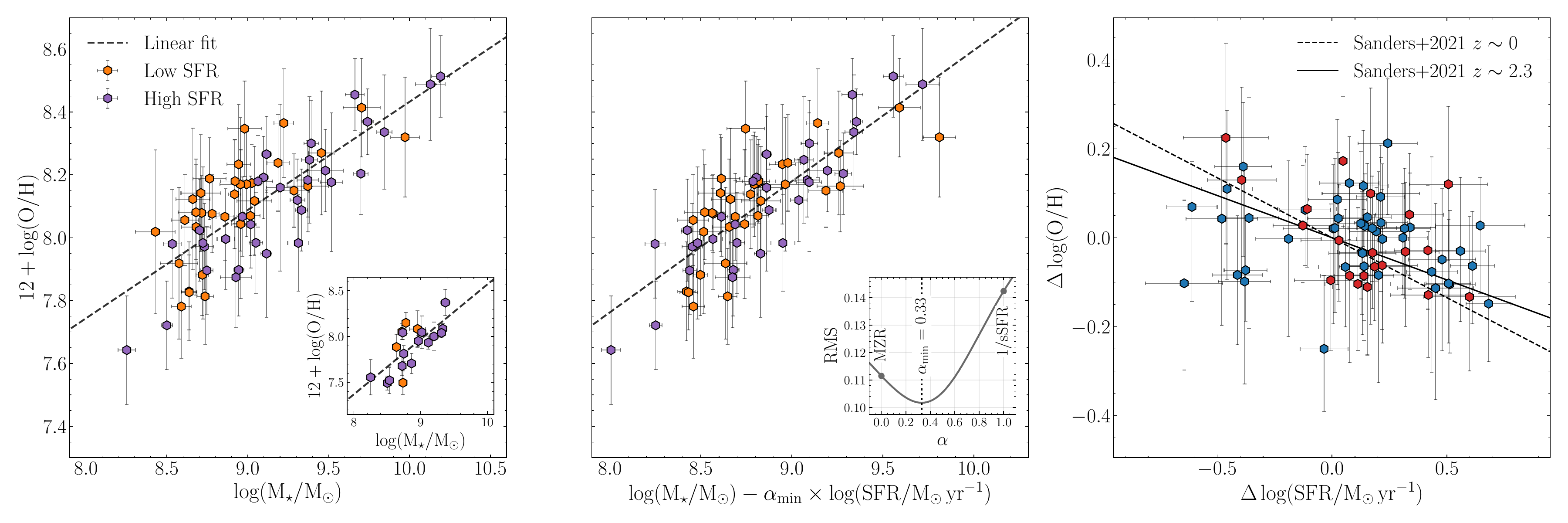}
            \caption{
            Parametric (left and centre panels) and non-parametric (right panel) tests for an FMR within the EXCELS sample.
            In the left panel, we show the mass-metallicity relationship of the EXCELS galaxies, with galaxies colour coded according to whether their SFR is less than (orange hexagons) or greater than (purple hexagons) the median SFR of the sample (i.e. $\log(\rm SFR/M_\odot\,yr^{-1})=0.71$).
            We plot a simple linear regression to the data with a black dashed line.
            In the inset panel, we show the the galaxies with direct-method metallicities along with the linear fit to the strong-line sample.
            The strong line metallicities show tentative evidence of a SFR-dependent scatter about the linear fit MZR, with low (high) SFR galaxies preferentially lying above (below) the MZR, consistent with an empirical FMR.
            Contrastingly, the direct-method metallicities show no clear signal of an SFR-dependent scatter.
            In the centre panel, we show best-fit parametric FMR assuming $\alpha=0.33$, with an inset panel shows the residual scatter as a function of different values of $\alpha$, and we note a $0.01\,\rm dex$ reduction in scatter for $\alpha=0.33$ versus the MZR.
            Whilst there is evidence of an SFR dependent scatter in the data, there is only a marginal reduction in scatter of $\sim10$ per cent.
            In the right panel, we show a non-parametric test for the FMR by comparing the residuals around the MZR, $\Delta\log(\rm O/H)$ against the residuals around the SFMS, $\Delta\log(\rm SFR)$.
            The low- and high-redshift subsamples are shown with blue and red hexagons, and we plot the trends from \citet{sanders2021} at $z=0, 2.3$ (dashed and solid lines).
            This test shows no conclusive evidence of a trend between $\Delta\log(\rm O/H)$ and $\Delta\log(\rm SFR)$ indicative of an FMR.
            }
            \label{fig:empirical-fmr}
        \end{figure*}

        To assess whether there is evidence for an FMR in our EXCELS sample we explore two of the standard approaches used in the literature. 
        First, we consider the parametric method presented in \citet{mannucci2010} where the parameter $\mu_\alpha$ is defined as a linear combination of stellar mass and star-formation rate:
        \begin{equation}
            \mu_\alpha = \log {\rm (M_\star/M_\odot)} - \alpha \log {\rm (SFR /M_\odot \, yr^{-1})}.
        \end{equation}
        In this expression, $\alpha$ introduces a SFR-dependent correction to the stellar mass, such that $\alpha=0$ corresponds to $\mu_0 = \mathrm{log}{\rm (M_\star/M_\odot)}$, and $\alpha=1$ returns the specific star-formation rate ($\mu_1 = \rm sSFR$).
        If there is an underlying FMR within our data, there will exist some value of $\alpha$ ($\alpha_{\rm min}$) that produces a $\mu_\alpha - \rm \log (O/H)$ relation which has less scatter than the MZR.
        In the local Universe, various studies have applied this approach and reported values of $0.2\leq\alpha_{\rm min}\leq0.7$ that minimise the scatter around the MZR by $\simeq20-70$ per cent \citep[e.g.][]{mannucci2010, andrews&martini2013, curti2020}

        We show the results of applying this test on our sample in Fig.~\ref{fig:empirical-fmr}.
        The left-hand panels shows the traditional MZR ($\alpha=0$) with the sample split by the median SFR.
        By eye, it can be seen that there is some tentative evidence for the FMR signature, with the higher SFR half of the sample to have lower metallicities, including in the direct-method metallicity subsample (inset panel).
        However, by exploring the scatter as a function of $\alpha$ we find that the scatter is not significantly reduced for any value of $\alpha$.
        The middle panels shows the projection of minimum scatter which has $\alpha_{\rm min}=0.33$ (similar to the value derived for the original FMR in \citealp{mannucci2010}).
        However, although the scatter is formally lower for this projection, the reduction in the scatter is $< 10$ per cent; we find $\sigma=0.102$ for $\alpha=0.33$ versus $\sigma=0.111$ for $\alpha=0$.
        This is significantly lower than reduction in scatter reported for local Universe FMR \citep[$\simeq20-70$ per cent,][]{yates2012, andrews&martini2013, curti2020}.
        We find an even smaller reduction in scatter using just our direct-method metallicity subset, with $\alpha_{\rm min}=0.16$ and $\sigma=0.186$ for this projection versus $\sigma=0.168$ for $\alpha=0$ (i.e. a $\lesssim2$ per cent reduction).
        Therefore, although we formally find a value of $\alpha$ which reduces the scatter, the effect is much smaller than that observed at $z=0$ and cannot be considered robust evidence for an FMR signature in our sample.
        
        Following the methodology introduced by \citet{salim2014}, we also perform a non-parametric test of FMR in which the offset from the MZR, $\Delta \log (\rm O/H)$, is correlated with the offset from the SFMS, $\Delta \log (\rm SFR)$.
        We show this for our low- and high-redshift bins in the right-hand panel of Fig.~\ref{fig:empirical-fmr}, where for each bin we take the residuals with respect to our best-fit MZR (Fig. \ref{fig:mass-metallicity-relationship}) and the \citet{clarke2024} SFMS (Fig. \ref{fig:star-forming-main-sequence}).
        Any correlation between these two variables would imply a SFR-dependence in the scatter around the MZR, and thus evidence an FMR.
        For example, \citet{sanders2021} find consistent negative correlations between $\Delta \log (\rm O/H)$ and $\Delta \log (\rm SFR)$ at $z=0$ and $z=2.3$ which we show in Fig. \ref{fig:empirical-fmr}.
        In our $2<z<4$ redshift bin, we do not see evidence of a correlation, and, although there is marginal evidence of of a negative correlation at $4<z<8$, the data are still clearly consistent with a flat relation.
        Applying a Spearman rank correlation test we indeed find that neither correlation is statistically significant (with Spearman coefficients of $\rho=[-0.10, -0.19]$ and $p=[0.44, 0.38]$ for $2<z<4$ and $4<z<8$ respectively).
        We discuss the application of this test to our direct-method sample in Appendix~\ref{sec:direct-method-non-parametric}.

    \subsection{A comparison with the locally-derived FMR} \label{sec:strong-line-fmr}

        \begin{figure*}
            \centering
            \includegraphics[width=\linewidth]{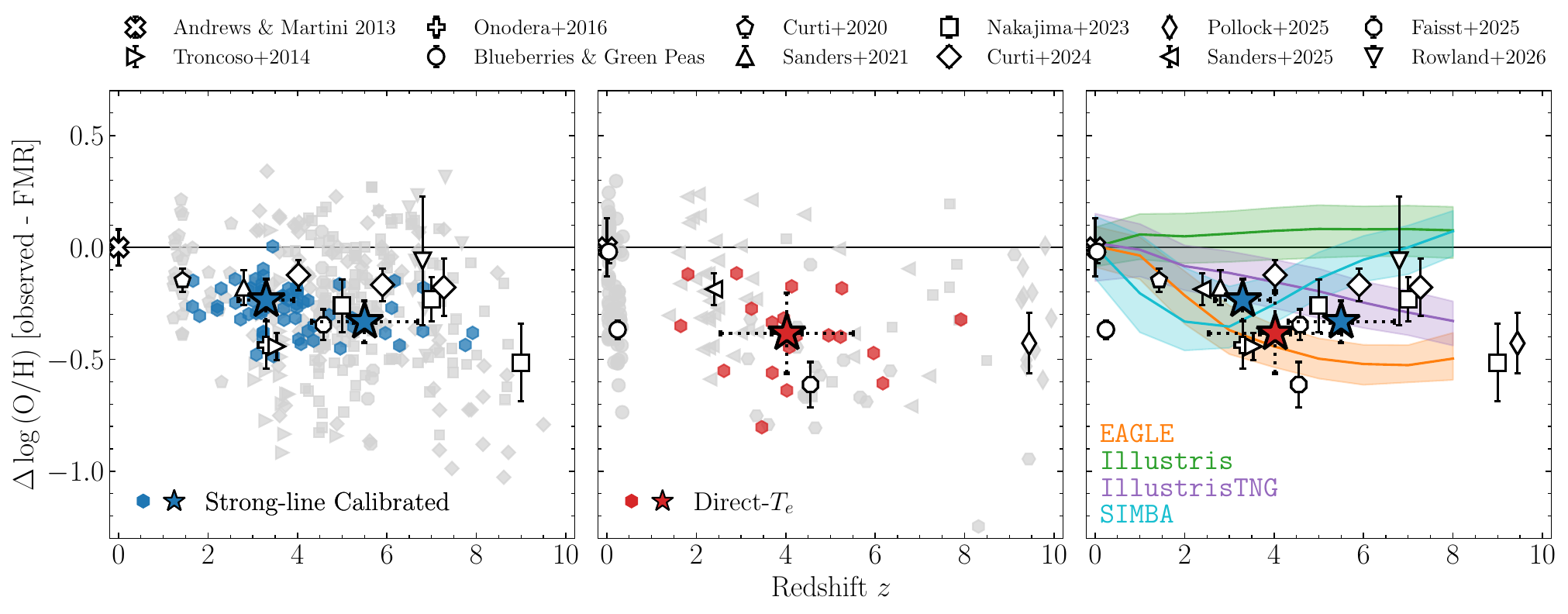}
            \caption{
            The deviation of our measured metallicities from the locally derived FMR from \citet{andrews&martini2013} (assuming $\alpha=0.66$) as a function of redshift.
            In the left panel, we show the results using strong-line derived metallicities (blue hexagons) alongside binned measurements corresponding to $2<z<4$ and $4<z<8$ (blue stars).
            Inverse-variance weighted uncertainties are shown by black error bars, and are typically smaller than the markers, so for clarity we also plot the dispersions for each bin with a dotted error bar.
            We show the results of \citet{troncoso2014, onodera2016, curti2020, curti2020b, sanders2021, rowland2025, heintz2023, nakajima2023, curti2024} and \citet{faisst2025a}, assigning grey symbols to individual galaxies and black and white markers to stacked or binned measurements. 
            In the centre panel, we show the results using strong-line derived metallicities (red hexagons) along with an average  measurement corresponding to $2<z<8$ (red star).
            As in the left panel, we plot the direct-method results from \citet{yang2017a, yang2017b, nakajima2023, faisst2025a} and \citet{pollock2025}.
            Both methodologies show deviation from the local FMR.
            In the right panel, we show the predictions of cosmological simulations with respect to their own $z=0$ FMR \citep[][]{garcia2024a, garcia2024b, garcia2025}, alongside all the stacks and binned measurements.
            The consensus between the majority of observations and simulations, including our results, favours an offset from the FMR that tentatively increases with redshift.
            }
            \label{fig:fmr-deviation}
        \end{figure*}

        In addition to looking for evidence for an FMR signature within high redshift samples, another question is whether these galaxies follow the form of the FMR derived for local galaxies.
        For a truly non-evolving FMR, galaxies at high redshift would be consistent with the locally-defined relation.
        In Fig.~\ref{fig:fmr-deviation}, we show the offsets of our galaxies (using both strong-line calibrated and auroral-line metallicities) from the local FMR calibration of \cite{andrews&martini2013} assuming $\alpha=0.32$.
        In addition showing individual galaxies, we also show averages for the strong-line sample at $2<z<4$ and $4<z<8$, and the average for our auroral-line sample across the full $2<z<8$ range.
        We additionally include an extensive set of literature results from both \emph{JWST} and pre-\emph{JWST} studies across the redshift range $0 < z < 10$. 
        These literature samples are drawn from the following surveys: SDSS \citet{curti2020}, KLEVER \citep[$z\sim1-2$;][]{curti2020b}, zCOSMOS \citep[$z\sim3.3$;][]{onodera2016}, ALPINE-CRISTAL-JWST \citep[$z\sim4-6$;][]{faisst2025a, faisst2025b}, REBELS \citep[$z\sim6-8$;][]{rowland2025}, and AURORA \citep[$z\sim1-7$;][]{sanders2025b}.
        We also include results for a number of individual systems from \citet{yang2017a, yang2017b} (local Blueberries and Green Peas), \citet{troncoso2014}, \citet{curti2023}, \citet{heintz2023}, \citet{morishita2024}, \citet{cullen2025} and \citet{pollock2025}.
        Where convenient, we average samples of individual measurements to illustrate the results more clearly.
        Although we have selected just one parametrisation of the local FMR to discuss here, we present alternative prescriptions of the local FMR in Appendix~\ref{sec:alternative-fmr-calibrations}.
        The general results discussed below hold regardless of the exact form of the $z=0$ FMR chosen for comparison.

        It can be seen that the galaxies in our sample are preferentially offset to lower metallicity than predicted by the FMR by up to $\simeq-0.5\,\rm dex$.
        For the strong-line sample, the low- and high-redshift averages have inverse-variance weighted FMR offsets of $-0.23\pm0.03$ and $-0.33\pm0.03\,\rm dex$ respectively (with dispersions of $\simeq0.1\,\rm dex$).
        This is consistent with the literature sample, which shows a similar degrees of offset that gradually increases with increasing redshift.
        Considering all of the data, the individual galaxies show significant variation in their offset from the FMR in  the $\Delta\log({\rm O/H})-z$ plane, from $-1.0$ to $0.5\,\rm dex$.
        However, we find that the majority of individual measurements from EXCELS (both strong-line and auroral) fall below the FMR prediction.
        The auroral-line sample we find an with an inverse-variance weighted average offset of $-0.38\pm0.02\,\rm dex$ at $\langle z \rangle \simeq 4$, with a dispersion of $\simeq0.18\,\rm dex$.
        Overall, our results are fully consistent with the previous literature reports of an increasing deviation from the $z=0$ FMR at $z \gtrsim 3$.

        In the right panel of Fig.~\ref{fig:fmr-deviation}, we compare our results and other observational literature results with the predictions of cosmological simulations as compiled by \citet{garcia2024b, garcia2025}.
        They find that the {\sc EAGLE}, {\sc SIMBA}, and {\sc IllustrisTNG} simulations each had a statistically significant evolution in $\alpha_{\rm min}$, dubbed a `weak' FMR, in which the relationship between the offset from the MZR and SFR changes with time.
        The existence of a `weak'-FMR suggests subtle changes in the operation of baryon cycle with redshift.
        The only simulation to not display an evolving FMR is {\sc Illustris}, which remains formally consistent with an FMR out to $z\simeq8$, while both {\sc IllustrisTNG} and {\sc EAGLE} show increasingly negative offsets from their $z=0$ FMRs with increasing redshift.
        In {\sc EAGLE}, the offset rapidly increases up to $z=2$ before plateauing, while {\sc IllustrisTNG} again agrees remarkably well with the observed trend.

        Overall, we find that our results are inconsistent with the local parametrisation of the FMR (see also Appendix~\ref{sec:alternative-fmr-calibrations}).
        The exact redshift at which the breakdown occurs is difficult to ascertain, but our results are consistent with this FMR breakdown occurring somewhere between $2<z<4$, consistent with previous studies \citep[e.g.][]{curti2023, curti2024, scholte2025, korhonen_cuestas2025}.
        We note, however, that our results remain consistent with the predictions of a weakly evolving FMR, and once again find the best agreement with the weakly-evolving FMR found in {\sc IllustrisTNG} \citep[][]{garcia2024b, garcia2025}. 
        Indeed, looking internally within our sample, we find tentative evidence for an FMR signature (Fig. \ref{fig:empirical-fmr}), but cannot robustly confirm it with the current data.
        Definitive confirmation of the existence or absence of the FMR at $z>3$ requires larger samples spanning wider ranges of stellar mass and SFR combined with, ideally, direct-method metallicities; such a dataset is achievable with future deep \emph{JWST} surveys.

    \subsection{The breakdown of the locally-derived FMR at \texorpdfstring{$\mathbf{z \gtrsim 3}$}{z>3}}

    In this section, we consider potential explanations for the breakdown of the locally-derived FMR in our sample and the other $z \gtrsim 3$ samples from the literature.
    We first consider whether how informative comparisons to the $z=0$ FMR are given the differences in the physical properties of the local and high-redshift samples.
    We then discuss some the physical explanations for an evolving relationship between mass, metallicity and star-formation rate.

    \begin{figure}
        \centering
        \includegraphics[width=1.0\linewidth]{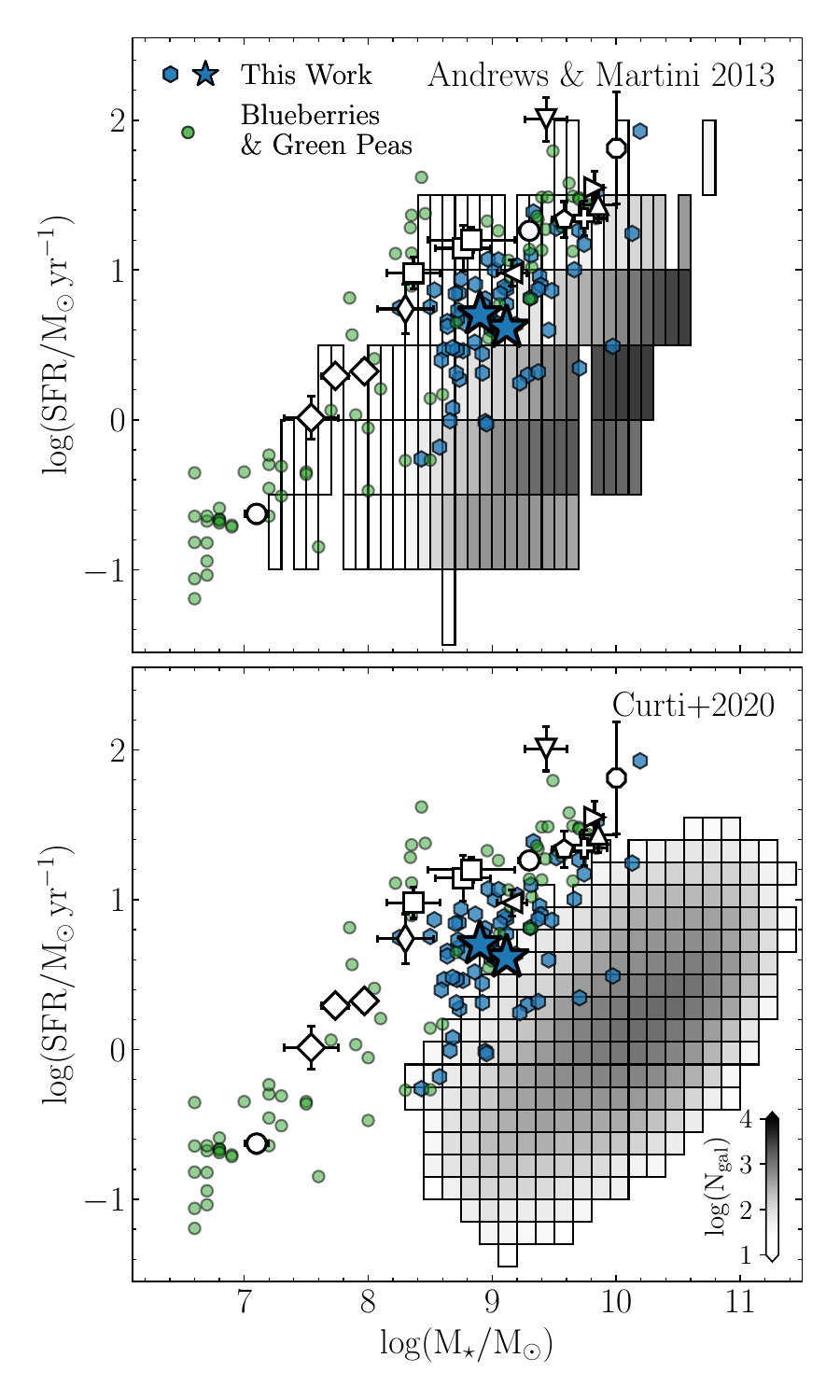}
        \caption{
        The local calibration samples (grey histograms) for the \citet{andrews&martini2013} (upper panel) and \citet{curti2020} FMRs (lower panel) in the star-forming main sequence plane.
        The individual EXCELS galaxies and binned measurements for the strong-line metallicities are plotted with blue hexagons and stars respectively.
        We plot the literature bins and stacks adopting the same symbols as Fig.~\ref{fig:fmr-deviation}, and plot individual blueberry and green pea galaxies from \citet{yang2017a, yang2017b} as small green circles.
        When accounting for the numbers of galaxies per bin, neither FMR parametrisation shows a large degree of overlap with the high-redshift samples, suggesting the observed offsets from the FMR could be due to differing physical properties (i.e. higher $\rm sSFR$).
        }
        \label{fig:fmr-calibration}
    \end{figure}

    \subsubsection{Extrapolating the local FMRs to high-redshift galaxies} \label{sec:fmr-calibration}

        It is first worth considering whether the samples used to define the FMR locally are representative of high-redshift galaxies in terms of their global properties.
        The properties of the galaxies considered `typical' at any redshift evolves, as reflected by the evolution in the SFMS \citep[i.e. higher SFR at fixed stellar mass with increasing redshift, e.g.][]{speagle2014, clarke2024} and the MZR \citep[i.e. lower metallicity at fixed stellar mass with increasing redshift, e.g.][]{tremonti2004, sanders2021}.
        It is therefore plausible that the offset from the FMR observed at high redshifts is in fact a product of extrapolating a locally derived relationship to high-redshift galaxies with differing physical properties (i.e. higher sSFRs).

        In  Fig.~\ref{fig:fmr-calibration} we show the overlap between the high redshift samples and two samples used to define the FMR locally \citep{andrews&martini2013, curti2020} in the stellar mass versus star-formation rate plane.
        It is immediately evident from Fig.~\ref{fig:fmr-calibration} that the local FMR is not well-calibrated for the high-sSFR galaxies commonly observed at high redshift. 
        Considering the \cite{andrews&martini2013} FMR used for Fig.~\ref{fig:fmr-deviation}, it can be seen that the vast majority of that sample has $\mathrm{log}(M_{\star}/\mathrm{M}_{\odot}) > 9.0$ and $\mathrm{log}(\mathrm{SFR}/\mathrm{M}_{\odot}\mathrm{yr}^{-1}) < 0.5$, significantly offset from the typical values of our EXCELS galaxies the other $z>3$ literature samples.
        Although there some galaxies in the \cite{andrews&martini2013} occupying a similar parameter space, the numbers of objects in these bins accounts for an extremely small fraction of the full sample (note a logarithmic scale is used for the number count shading in Fig.~\ref{fig:fmr-calibration}).
        The same is true for the \citet{curti2020} $z=0$ FMR sample.
        It is clear that neither $z=0$ FMR is well constrained in the parameter space occupied by high-redshift sources.

        Notably, it can also be seen that the same is true for the $z\simeq0$ `Blueberry' and `Green Pea' galaxies \citep{yang2017a, yang2017b}.
        These galaxies show an offset below the \citet{andrews&martini2013} FMR by as much as $0.5\,\rm dex$, in line with the $z>3$ samples (Fig.~\ref{fig:fmr-deviation}). 
        From Fig.~\ref{fig:fmr-calibration} it can also be seen that they have similar stellar masses and star-formation rate to the $z>3$ samples.
        Indeed, these galaxies are known to be young, highly ionized and strongly star-forming systems, and are often considered analogues of high-redshift galaxies.
        This comparison strongly suggests that the FMR offset is primarily a function of galaxy properties, rather than redshift.

        In fact, this point has recently been conclusively demonstrated by \citet{laseter2025}, who investigated the FMR for $\simeq 700$ galaxies at $z=0$ with $\log ({\rm M_\star/M_\odot}) < 9$, finding no evidence for an FMR in these low-mass local systems.
        It is therefore possible that the baryon cycle equilibrium that gives rise to an FMR signature in $\log ({\rm M_\star/M_\odot}) > 9$ galaxies at $z=0-2$ does not apply to galaxies below that mass threshold at any redshift.
        In this scenario, the increasingly prominent offset from the FMR with redshift would be a result of the average population shifting to lower mass and higher sSFR.
        On that point, it is interesting note that high-redshift studies specifically targeting the most massive galaxies at $z\simeq4-7$ have found better consistency with the local FMR \citep[e.g.][]{faisst2025a, rowland2025}.
        
        To summarise, for our EXCELS sample, which has $\langle \log ({\rm M_\star/M_\odot}) \rangle \simeq 9.0$, we clearly find an offset from the $z=0$ FMR consistent with previous high-redshift studies.
        However, a simple redshift-evolution interpretation is deceptive, since the local FMR was not calibrated for galaxies with similar stellar masses and star-formation rates.
        Local galaxies matched in these properties also show a similar offset from the canonical $z=0$ FMR \citep[e.g.][]{yang2017a, yang2017b, laseter2025}.
        That being said, we can't yet rule out some form of SFR-dependent scatter around the MZR within our sample (Fig.~\ref{fig:fmr-deviation}).
        Again, large, homogeneously analysed samples at $z>3$ are still needed to fully understand the link between mass, metallicity and star-formation at these cosmic epochs.

    \subsubsection{What drives the FMR offset of high-redshift galaxies?}

        The physical mechanism, or mechanisms, driving the breakdown of the FMR at $z \gtrsim 3$, as well as in local low-mass galaxies, have been subject to significant discussion in the recent literature.
        One commonly proposed solution relates to the large gas fractions that are a known feature of the high-sSFRs galaxies found at high redshifts \citep{tacconi2018}.
        In this picture, significant `pristine' gas inflows will dilute the gas-phase metallicity while providing ample material to boost star formation, potentially causing significant deviations from an equilibrium FMR \citep{heintz2023, li2025, pollock2025}.
        Indeed, some galaxies in the early Universe have been shown to contain strong damped Ly\,$\alpha$ absorption associated with their ISM, indicative of large pristine gas fractions within galaxies at this epoch \citep[e.g.][]{umeda2024, heintz2025}.
        There is also evidence that $z \simeq 3$ represents a critical epoch when gas inflows into galaxies transitioned from predominantly pristine to enriched \citep[e.g.][]{møller2013, heintz2022}.

        However, it is not clear that this explanation can account for the similar FMR offsets observed in low-mass, high-sSFR $z=0$ galaxies.
        For example, \citet{laseter2024} find no correlation between gas fraction and FMR deviation in their low-mass local sample, while at the same time observing the expected trend between high gas-fraction and elevated SFR.
        Instead, they find that the metal content of the ISM relative to the overall gas fraction (the `effective yield') is directly correlated with the offset from the local FMR.
        They argue that, rather than pristine gas inflows, star-formation and enriched outflows in bursty systems out of gas reservoir equilibrium are driving the FMR offsets in these systems.
        This could of course also apply to $z>3$ galaxies, since formation histories in the early Universe have been shown to become increasingly stochastic (or `bursty'; e.g. \citealp{sun2023, endsley2025}).
        Indeed, some simulations also suggest that the competition between short-timescale variations in SFR and enrichment timescales are a factor in driving the offset in the FMR \citep[e.g.][]{torrey2018, marszewski2025}.

        Of course, it is possible that local low-mass systems and $z>3$ galaxies might show FMR offsets for different reasons.
        For example, while for the $z=0$ sample of \citet{laseter2025} the MZR scatter is not SFR-dependent, there are tentative hints at a SFR dependence in our $2 < z < 8$ sample, which would hint at different physical processes at play.  
        Definitively determining whether or not the MZR scatter shows any dependence on SFR in the $z>3$ population is one clear way to make further progress here.

\section{Summary \& Conclusions} \label{sec:conclusions}

    In this work, we have presented a homogenous analysis of $65$ star-forming galaxies from the EXCELS survey spanning the redshift range $2 < z < 8$ to explore the relationship between mass, metallicity and star-formation rate at these redshifts.
    We estimate gas-phase metallicities (i.e. oxygen abundance) from strong nebular emission lines using the strong-line calibration scheme described in \citet{scholte2025}, which has been explicitly shown to recover the metallicities of EXCELS galaxies.
    Additionally, we measure direct metallicities based on the \oiiib \ line for $19$ galaxies in our sample.
    The stellar masses are estimated from SED fitting using the deep \emph{HST} and \emph{JWST}/NIRCam photometric coverage for all galaxies in our sample, and the SFRs derived from Balmer line emission.
    Our main results can be summarised as follows:
    
    \begin{enumerate}
    
        \item The gas-phase metallicities across our full sample range between $\simeq 0.1-0.7 \, \rm Z_\odot$ across the stellar masses range $10^8-10^{10.5}\,\rm M_\odot$, with a clear correlation between metallicity and stellar mass (i.e. the MZR; Fig. \ref{fig:mass-metallicity-relationship}).
        Splitting our sample into two redshift bins ($2 < z < 4$ and $4 < z < 8$), we find that the gradient of the MZR remains constant with a slope of $\rm d\log(O/H)/d \log (M_\star)\simeq0.29\pm0.01$ and $-0.30\pm0.02$ at $\langle z \rangle \simeq3.3$ and $\langle z \rangle \simeq 5.5$ respectively (Fig. \ref{fig:mass-metallicity-relationship}).
        However, the normalisation of the MZR at $10^{9}\,\rm M_\odot$ deceases from $Z_{9} = 8.14\pm0.01$ at $\langle z \rangle \simeq3.3$  to $Z_{\rm 9}=8.00\pm0.02$ at $\langle z \rangle \simeq 5.5$ consistent with the expected evolution towards lower metallicities at earlier cosmic epochs.

        \item We find that the auroral-line subsample is fully consistent with the relations derived from the strong lines, indicating that reliable metallicity scaling relation can be inferred using strong-line ratios at $z>3$ as long as appropriate calibrations are applied (Fig.~\ref{fig:mass-metallicity-relationship}).
        However, we find evidence for an artificial reduction in the MZR scatter when applying strong line ratios, highlighting the fact that these calibration do not perfectly capture the variation in ionization conditions at fixed metallicity.

        \item We compile a collection of literature MZRs to investigate the overall evolution of the MZR slope and normalisation across $0 < z < 10$ (Figs.~\ref{fig:mzr-lit-sim-comparison} and ~\ref{fig:mzr-parameter-evolution}).
        Combined, these observations suggest that the slope of the MZR does not evolve strongly with redshift but the normalisation clearly decreases with increasing redshift.
        Comparing these results with the predictions of cosmological simulations, we find a fully consistent picture (Fig.~\ref{fig:mzr-parameter-evolution}), with {\sc IllustrisTNG} best reproducing observational trends.

        \item We find that our new MZR at $\langle z \rangle \simeq 5.5$ is fully consistent with a redshift extrapolation of the relation presented in \citet{jain2025}, which had previously been calibrated out to $z\simeq3$ (Fig.~\ref{fig:evolving-mzr}). 
        In this prescription, the redshift evolution of the MZR can be captured with one parameter: the redshift-dependent turnover mass.
        
        \item Our results imply that galaxies at $z=5.5$ $(z=3.2)$ with $M_{\star} = 10^9 \, \mathrm{M}_{\odot}$ are already enriched at $\simeq30$ per cent ($\simeq40$) per cent of the oxygen abundance observed in equivalent stellar mass at $z=0$ by (Fig.~\ref{fig:cosmic-metallicity-evolution}).
        Combined with the other literature measurements, these results are consistent with rapid gas-phase metallicity enrichment within the first $15$ per cent of cosmic time (i.e. the first $2 \, \rm{Gyr}$).

        \item We investigate evidence for an FMR signature in our sample and find tentative, but inconclusive evidence, of a SFR-dependent scatter around the MZR (Fig.~\ref{fig:empirical-fmr}).
        Applying the parametric approach introduced by \citet{mannucci2010}, we find an optimal FMR projection with $\alpha=0.33$; however, the reduction in scatter relative to the standard MZR is only $\simeq0.01$ dex, much less than the $\simeq0.05-0.1$ dex reduction in scatter seen at $z=0$ \citep[e.g.][]{mannucci2010, andrews&martini2013}.
        Applying a non-parametric approach \citep[e.g.][]{salim2014} we similarly see a weak, but not significant, correlation between the degree of offset from the star-formation main sequence and offset from the MZR (Fig.~\ref{fig:empirical-fmr}).
        We find similar results when considering just the auroral line sample.
        
        \item Comparing our sample directly to parametrisations of the $z=0$ FMR, we find a clear offset consistent with that reported by other studies at $z>3$ (Fig. \ref{fig:fmr-deviation}).
        However, it not clear that this is a purely redshift-drive effect.
        For example, we note similar offsets to local `Blueberry' and `Green pea' systems, which are analogues of high-redshift galaxies.
        Indeed, comparing to the masses and star-formation rates of the \citet{andrews&martini2013} and \citet{curti2020} FMR samples, we find both the $z>3$ galaxies and their $z=0$ analogues exhibit minimal overlap, being offset to much higher sSFRs (Fig.~\ref{fig:fmr-calibration}).
        Our inference is similar to that presented in \citet{laseter2025}, who argue that the local FMR - calibrated using galaxies with $\mathrm{log}(M_{\star}/\mathrm{M}_{\odot}) > 9.0$ - is not applicable to lower-mass galaxies even at $z=0$.
        However, it remains unclear whether the physical reasons for the FMR offset is the same for low-mass $z=0$ galaxies and galaxies at $z \gtrsim 3$.
        
    \end{enumerate}

    Our analysis highlights the capability of \emph{JWST} surveys such as EXCELS to constrain the oxygen abundances and physical properties of star-forming galaxies over a large span of cosmic time.
    Future \emph{JWST} surveys aimed at securing large numbers of auroral line detections will be essential to better constrain the MZR, and assess the existence of the FMR, at $z>3$.

\section*{Acknowledgements}

    TMS, FC, KZA-C and DS acknowledge support from a UKRI Frontier Research Guarantee Grant (PI Cullen; grant reference: EP/X021025/1). 
    ACC, H-HL, ET and SDS acknowledge support from a UKRI Frontier Research Guarantee Grant (PI Carnall, grant reference EP/Y037065/1). 
    RB, CB, and RJM acknowledge the support of the Science and Technology Facilities Council. JSD and DJM acknowledge the support of the Royal Society through the award of a Royal Society University Research Professorship to JSD.

    The following software was utilised: {\sc astropy} \citep{astropy_collaboration2022}, {\sc bagpipes} \citep{carnall2018, carnall2019}, {\sc Dynesty} \citep{speagle2020, koposov2023}, {\sc matplotlib} \citep{hunter2007}, {\sc numpy} \citep{harris2020}, {\sc PyNeb} \citep{luridiana2015, morisset+2020} and {\sc scipy} \citep{virtanen2020}.
    This research utilised NASA's Astrophysics Data System Bibliographic Services.
    
    For the purpose of open access, the author has applied a Creative Commons Attribution (CC BY) licence to any Author Accepted Manuscript version arising from this submission. 

\section*{Data Availability}

    All \textit{JWST} and HST data products are available via the Mikulski Archive for Space Telescopes (\url{https://mast.stsci.edu}). 
    The derived data products presented in Tables~\ref{tab:sample-properties} and \ref{tab:direct-metallicities} are available in machine readable format (\href{https://tmstanton.github.io/publications}{https://tmstanton.github.io/publications}).
    Additional data products are available from the authors upon reasonable request.    
    

\bibliographystyle{mnras}
\bibliography{strong-lines.bib} 


\appendix

    \section{Additional AGN diagnostic diagrams} \label{sec:oiii4363-diagnostics}

        \begin{figure*}
            \centering
            \includegraphics[width=0.95\linewidth]{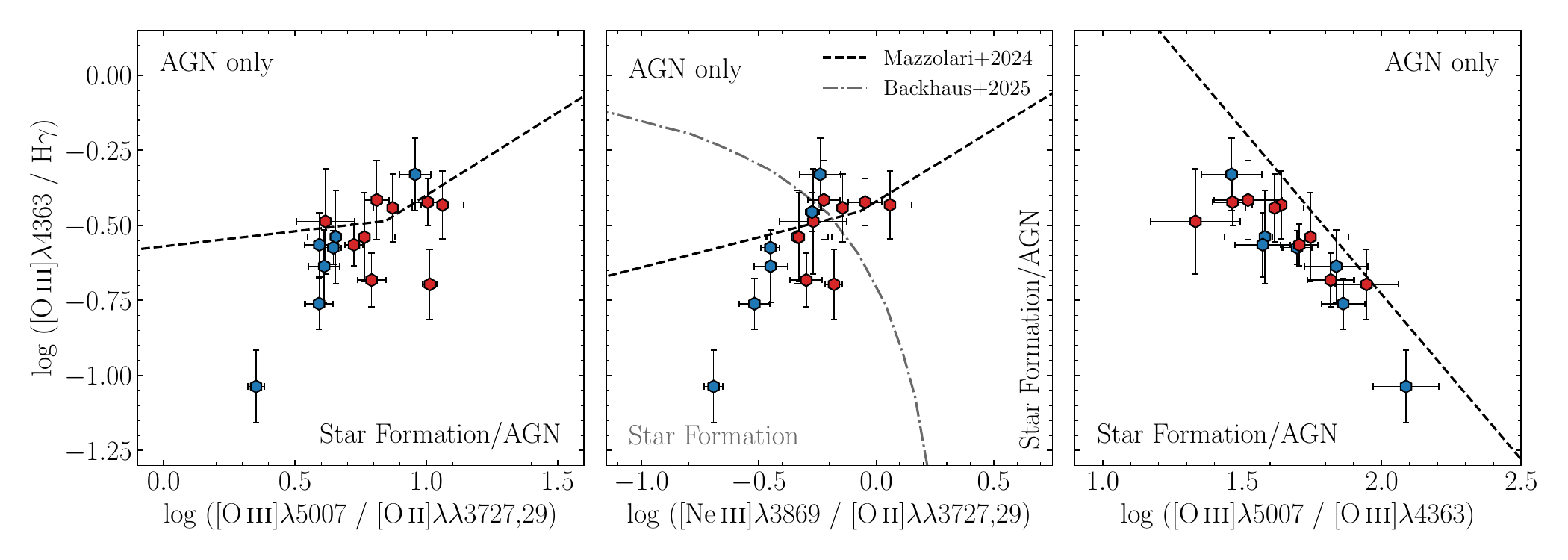}
            \caption{
            Additional diagnostic diagrams for AGN using the \oiiib/\hgamma \ ratio versus the \oiiia/\oii \ ratio (left panel), \neiii/\oii \ ratio (centre panel), and \oiiia/\oiiib \ ratio (right panel).
            Separations between regions of the diagrams corresponding to AGN-only sources and star-formation or AGN sources from \citet{mazzolari2024} and \citet{backhaus2025} are shown in black-dashed and grey dot-dashed lines respectively.
            The EXCELS galaxies (blue hexagons, $z<4$; red hexagons, $z>4$) with significant ($\rm SNR>3$) detections of the relevant emission lines are all $1\sigma$ consistent with the `Star Formation or AGN' regions of the three diagrams, indicating no strong evidence for AGN contamination within our sample.
            }
            \label{fig:oiii4363-diagnostic-diagrams}
        \end{figure*}

        As an additional test for narrow-line AGN within our sample, in Fig.~\ref{fig:oiii4363-diagnostic-diagrams} we place our galaxies on the \oiiib/\hgamma \ diagnostic diagrams presented by \citet{mazzolari2024} and \citet{backhaus2025}.
        All of our galaxies are consistent with the `star formation or AGN' regions of the \oiiib/\hgamma \ ratio versus the \oiiia/\oii \ and \oiiib/\hgamma \ ratio versus the \oiiia/\oiiib \ diagrams.
        Although one galaxy in the \oiiib/\hgamma \ versus \neiii/\oii \ diagram falls in the `AGN-only' region according to the \citet{mazzolari2024} separation line, this galaxy is consistent with star formation region defined by \citet{backhaus2025}.
        Overall, we find no strong evidence for AGN contamination within our sample using these diagnostics.

    \section{The direct-method metallicities of EXCELS galaxies} \label{sec:direct-method-appendix}

        \begin{table*} 
\caption{Direct-$T_e$ measurements of the full sample compared with their reported literature measurements in \citet{scholte2025}.}
\label{tab:direct-metallicities}
\renewcommand{\arraystretch}{1.3}
\begin{tabularx}{\linewidth}{@{\extracolsep{\fill}}ccccccc@{}} 
\toprule
ID & $T_e$ ([O {\sc iii}]) & $n_e$ ([S {\sc ii}])  & ${\rm O^+/H^+}$ & ${\rm O^{++}/H^+}$ & \multicolumn{2}{c}{$12 + {\rm \log (O/H)}$}  \\ 
   & / K & / cm$^{-3}$ & / $10^{-5}$ & / $10^{-5}$ & \multicolumn{1}{c}{(This Work)} & \multicolumn{1}{c}{(\citealp{scholte2025})} \\ 
\midrule \midrule
40081 & $12900^{\,+600}_{-600}$ & $226^{\,+103}_{-91}$ & $2.55^{\,+0.40}_{-0.31}$ & $9.62^{\,+1.39}_{-1.13}$ & $8.09^{\,+0.06}_{-0.05}$  & $7.98^{\,+0.06}_{-0.06}$  \\
41292 & $17300^{\,+1800}_{-1800}$ & $135^{\,+623}_{-111}$ & $0.94^{\,+0.38}_{-0.23}$ & $4.12^{\,+1.24}_{-0.81}$ & $7.71^{\,+0.12}_{-0.10}$ & --- \\
45177 & $12900^{\,+600}_{-600}$ & $479^{\,+374}_{-243}$ & $2.90^{\,+1.06}_{-0.64}$ & $11.23^{\,+3.97}_{-2.28}$ & $8.15^{\,+0.13}_{-0.10}$ &  $8.08^{\,+0.12}_{-0.10}$  \\
45393 & $13400^{\,+1600}_{-1700}$ & --- & $3.25^{\,+2.15}_{-1.01}$ & $6.69^{\,+3.20}_{-1.70}$ & $8.00^{\,+0.18}_{-0.14}$ &  $7.98^{\,+0.12}_{-0.10}$  \\
47557 & $13500^{\,+1800}_{-2000}$ & $90^{\,+433}_{-69}$ & $2.11^{\,+1.41}_{-0.64}$ & $8.99^{\,+5.52}_{-2.50}$ & $8.05^{\,+0.21}_{-0.14}$  & $7.92^{\,+0.16}_{-0.13}$  \\
52422 & $20200^{\,+1800}_{-1600}$ & --- & $0.28^{\,+0.12}_{-0.06}$ & $2.81^{\,+0.54}_{-0.44}$ & $7.49^{\,+0.08}_{-0.08}$ &  $7.47^{\,+0.10}_{-0.09}$ \\
57498 & $15000^{\,+700}_{-700}$ & --- & $1.70^{\,+0.69}_{-0.27}$ & $7.10^{\,+0.98}_{-0.72}$ & $7.95^{\,+0.06}_{-0.05}$ &  $7.97^{\,+0.05}_{-0.04}$  \\
59009 & $13500^{\,+1800}_{-2000}$ & --- & $1.73^{\,+0.72}_{-0.35}$ & $9.20^{\,+1.67}_{-1.32}$ & $8.04^{\,+0.08}_{-0.07}$ &  $8.15^{\,+0.06}_{-0.06}$ \\
60713 & $22400^{\,+3800}_{-3500}$ & --- & $0.50^{\,+0.34}_{-0.17}$ & $2.81^{\,+1.20}_{-0.67}$ & $7.52^{\,+0.16}_{-0.12}$ & ---  \\
66899 & $18400^{\,+3800}_{-3400}$ & --- & $0.31^{\,+0.28}_{-0.13}$ & $3.27^{\,+2.07}_{-1.04}$ & $7.55^{\,+0.22}_{-0.17}$  & --- \\
70864 & $13800^{\,+1000}_{-1000}$ & --- & $1.23^{\,+0.71}_{-0.34}$ & $9.98^{\,+2.21}_{-1.69}$ & $8.05^{\,+0.09}_{-0.08}$  & $8.18^{\,+0.14}_{-0.11}$  \\
83628 & $20500^{\,+3100}_{-2700}$ & --- & $0.31^{\,+0.19}_{-0.10}$ & $2.80^{\,+0.97}_{-0.68}$ & $7.50^{\,+0.13}_{-0.12}$  & ---  \\
93897 & $15100^{\,+900}_{-900}$ & --- & $1.50^{\,+0.62}_{-0.29}$ & $6.99^{\,+1.24}_{-0.96}$ & $7.93^{\,+0.08}_{-0.07}$ & $8.19^{\,+0.10}_{-0.11}$  \\
94335 & $10200^{\,+900}_{-1000}$ & $424^{\,+191}_{-135}$ & $7.09^{\,+2.83}_{-1.75}$ & $16.46^{\,+7.65}_{-4.16}$ & $8.37^{\,+0.16}_{-0.12}$ &  $8.22^{\,+0.10}_{-0.08}$  \\
95839 & $14100^{\,+600}_{-600}$ & --- & $1.57^{\,+0.63}_{-0.25}$ & $9.23^{\,+1.22}_{-1.00}$ & $8.04^{\,+0.06}_{-0.05}$ &  $8.03^{\,+0.06}_{-0.05}$ \\
104937 & $13300^{\,+2000}_{-2200}$ & $38^{\,+91}_{-23}$ & $4.15^{\,+3.02}_{-1.38}$ & $7.94^{\,+5.49}_{-2.48}$ & $8.08^{\,+0.23}_{-0.17}$ &  $8.16^{\,+0.19}_{-0.14}$ \\
119504 & $16300^{\,+2000}_{-2100}$ & --- & $0.74^{\,+0.50}_{-0.24}$ & $6.91^{\,+2.91}_{-1.66}$ & $7.89^{\,+0.15}_{-0.12}$ & $7.58^{\,+0.17}_{-0.14}$  \\
121806 & $16600^{\,+1400}_{-1300}$ & $275^{\,+2427}_{-247}$ & $0.80^{\,+0.34}_{-0.19}$ & $5.65^{\,+1.27}_{-0.95}$ & $7.81^{\,+0.09}_{-0.08}$  & $7.93^{\,+0.09}_{-0.08}$  \\
123837 & $17700^{\,+1800}_{-1800}$ & --- & $1.01^{\,+0.54}_{-0.26}$ & $3.67^{\,+1.06}_{-0.69}$ & $7.68^{\,+0.11}_{-0.10}$ &  $7.68^{\,+0.12}_{-0.11}$ \\ \bottomrule 
\end{tabularx}
\end{table*}

        Our electron temperature, density, and direct-method metallicity estimates for 19 EXCELS galaxies are reported in Table~\ref{tab:direct-metallicities}, alongside the measurements from \citet{scholte2025}.
        The main difference between these calculations is the version of the EXCELS reductions; our calculations use the current reductions as described in Section~\ref{sec:data-reduction}, crucially including improved artefact removal and leverage of the full sample of EXCELS galaxies, whilst \citet{scholte2025} used the previous reductions and a subset of the full galaxy sample.
        Compared to \citet{scholte2025}, we report new direct-method metallicities for four galaxies (41292, 60713, 66899, and 83628).
        However, we are unable to recover metallicities for seven galaxies (48659, 56875, 59720, 63962, 69991, 73535, and 123597).
        For 56875, 59720, 63962, 69991, 73535, and 123597, the new reductions place the \oiiib \ emission line below the detection threshold of $\rm S/N>3$, and for 48659 the \oii \ doublet is now below the detection threshold.
        Our metallicity measurements are highly consistent with \citet{scholte2025}, with an average offset of $0.01\pm0.04\,\rm dex$, with only two galaxies showing discrepancies of $>1\sigma$ (93897 and 119504).

    \section{Applying the non-parametric test to direct-method metallicites} \label{sec:direct-method-non-parametric}

\begin{figure}
            \centering
            \includegraphics[width=0.95\linewidth]{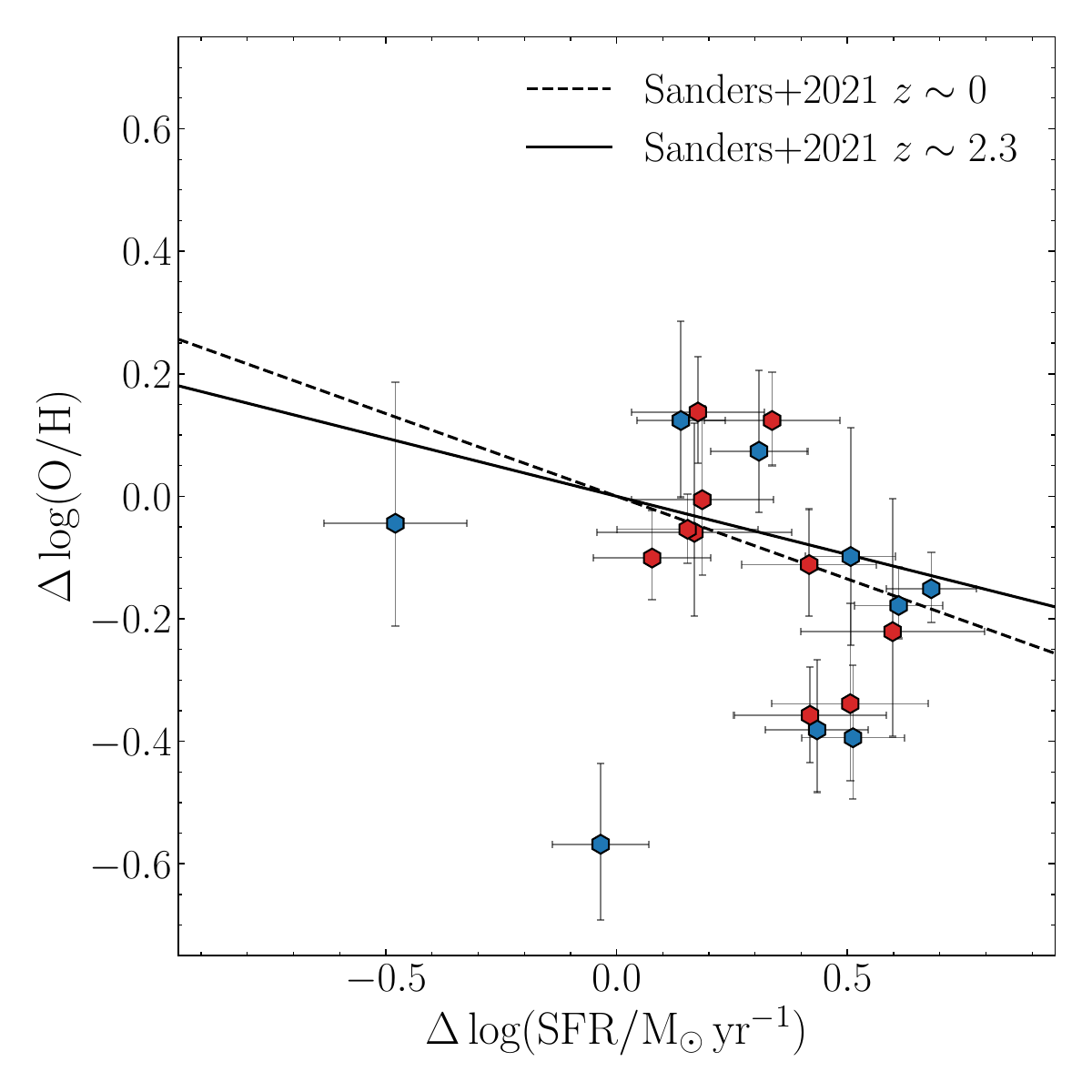}
            \caption{
            The non-parametric test for the FMR from Section~\ref{sec:strong-line-fmr-empirical} applied to the direct-method metallicity sample.
            We compare the residuals around the MZR, $\Delta\log(\rm O/H)$ to the residuals around the SFMS, $\Delta\log(\rm SFR)$.
            The low- and high-redshift subsamples are shown with blue and red hexagons, and we plot the trends from \citet{sanders2021} at $z=0, 2.3$ (dashed and solid lines).
            This test shows less scatter between $\Delta\log(\rm O/H)$ and $\Delta\log(\rm SFR)$ than the strong-line sample, and a more tentative signal of an FMR, highlighting the need for larger direct-method metallicity samples. 
            }
            \label{fig:direct-method-nonparam}
        \end{figure}

        In Fig.~\ref{fig:direct-method-nonparam} we perform the non-parametric test for the FMR \citep[][]{salim2014} on our direct-method metallicity sample.
        Whilst there correlations are not statistically significant, with Spearman coefficients of $\rho=[-0.13, -0.38]$ and $p=[0.64, 0.28]$ for $2<z<4$ and $4<z<8$ respectively, there remains a potential tentative signal of a correlation between $\Delta \log(\rm O/H)$ and $\Delta \log(\rm SFR)$.
        Notably, the EXCELS auroral line sample (as well as other auroral line samples, see \citealp{sanders2025b}) tend to be biased above the SFMS.
        This highlights the necessity for deep \emph{JWST} surveys to constrain direct method metallicities for samples spanning larger ranges of stellar mass and SFR.

    \section{Alternative local FMR calibrations} \label{sec:alternative-fmr-calibrations}

\begin{figure*}
            \centering
            \includegraphics[width=\linewidth]{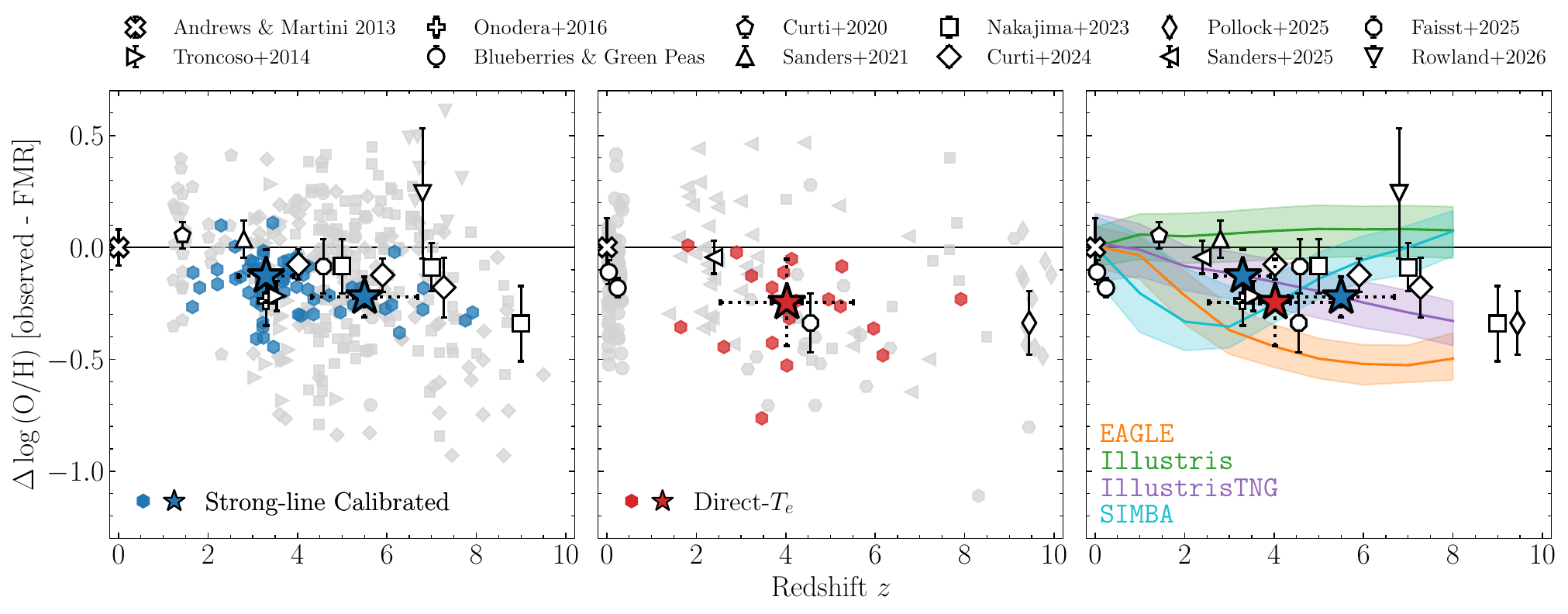}
            \caption{
            The deviation of our measured metallicities from the locally derived FMR from \citet{andrews&martini2013} (assuming $\alpha=0.32$) as a function of redshift.
            Points are labelled identically to Fig.~\ref{fig:fmr-deviation}. 
            }
            \label{fig:fmr-deviation-am13:0.32}
        \end{figure*}

\begin{figure*}
            \centering
            \includegraphics[width=\linewidth]{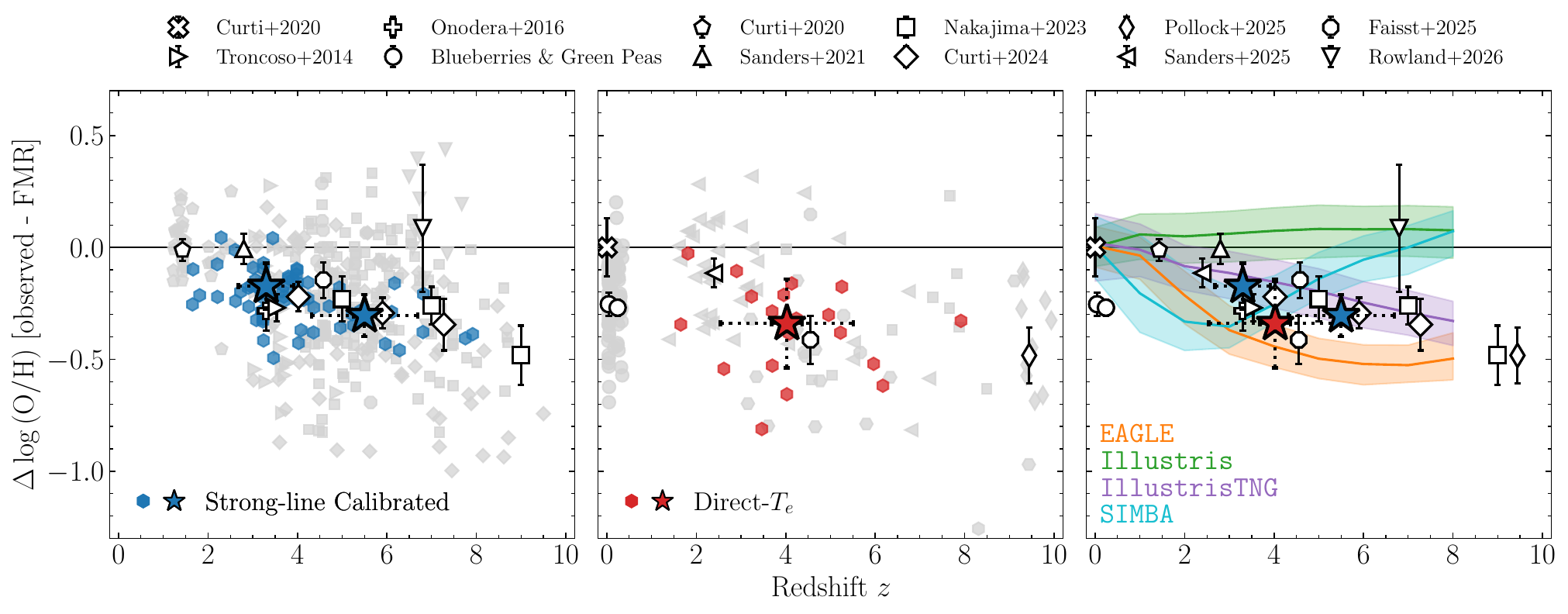}
            \caption{
            The deviation of our measured metallicities from the locally derived FMR from \citet{curti2020} (corresponding to $\alpha=0.55$) as a function of redshift.
            Points are labelled identically to Fig.~\ref{fig:fmr-deviation}.
            }
            \label{fig:fmr-deviation-c20}
        \end{figure*}

        In Figs.~\ref{fig:fmr-deviation-am13:0.32} and \ref{fig:fmr-deviation-c20}, we show alternative versions of Fig.~\ref{fig:fmr-deviation} applying the \citet{andrews&martini2013} local FMR (with $\alpha=0.32$) and the \citet{curti2020} FMR respectively.
        Our observations do not agree with the local FMRs in either case, though we see improved agreement with the offsets observed in the literature for the \citet{curti2020} FMR calibration.

        Assuming the \citet{andrews&martini2013} local FMR with $\alpha=0.66$, the average offsets for our strong-line calibrated metallicity bins are $-0.12\pm0.03$ and $-0.22\pm0.03\,\rm dex$ for $z\sim3.3, 5.5$ respectively, with corresponding dispersions of $\simeq0.10\,\rm dex$.
        The direct-method bin at $z\sim4$ is offset by $-0.25\pm0.02\,\rm dex$ with a dispersion of $\simeq0.19\,\rm dex$.
        If assuming the \citet{curti2020} local FMR, the average offsets for our strong-line calibrated metallicity bins are $-0.17\pm0.02$ and $-0.30\pm0.03\,\rm dex$ for $z\sim3.3, 5.5$ respectively, with corresponding dispersions of $\simeq0.10\,\rm dex$.
        For this FMR, the direct-method bin at $z\sim4$ is offset by $-0.34\pm0.02\,\rm dex$ with a dispersion of $\simeq0.20\,\rm dex$.

\bsp	
\label{lastpage}
\end{document}